\documentclass[12pt]{article}
\usepackage{a4wide}
\usepackage{epsfig}
\newcommand{\eps}{\varepsilon}
\newcommand{\slk}{/\kern-6pt k}
\newcommand{\sll}{/\kern-4pt l}
\newcommand{\slp}{p\kern-5pt/}
\newcommand{\slq}{q\kern-5.5pt/}

\newcommand{\oone}{\hbox{$1\kern-2.5pt\hbox{\rm l}$}}
\newcommand{\ssigma}{\hbox{$\kern2.5pt\vrule height4pt\kern-2.5pt\sigma$}}
\newcommand{\MeV}{{\rm\,MeV}}
\newcommand{\GeV}{{\rm\,GeV}}

\newcommand\pfrac[2]{\left(\frac{#1}{#2}\right)}
\newcommand{\Li}{{\rm Li}}
\newcommand{\real}{\mathop{\rm Re}\nolimits}

\newcommand{\sla}{{\sqrt\lambda}}

\newcommand{\slell}{/\kern-5pt\ell}
\newcommand{\arctanh}{\mathop{\rm arctanh}\nolimits}

\begin{document}

\thispagestyle{empty} 

\begin{center}
{\Large\bf Identical particle and lepton mass effects in the decay
$H\to Z^\ast(\to\tau^+\tau^-)+Z^\ast(\to\tau^+\tau^-)$}\\[1.3cm]
{\large  Stefan Groote\footnote{stefan.groote@ut.ee}, Lauri
Kaldam\"ae\footnote{kaldamae@protonmail.com} and Maria
Naeem\footnote{maria.naeem@ut.ee}}\\[1cm]
F\"u\"usika Instituut, Tartu \"Ulikool, W.~Ostwaldi~1, 50411 Tartu, Estonia
\end{center}

\vspace{1cm}
\begin{abstract}\noindent
We consider identical particle and lepton mass effects in the cascade decay
$H\to Z^\ast(\to\tau^+\tau^-)+Z^\ast(\to\tau^+\tau^-)$ and subordinate leading
order decays with the same final state. Since the scale of the problem is set
by the off-shellness $p_a^2$ and $p_b^2$ of the respective gauge bosons in the
limits ($4m_\tau^2\le p_a^2,p_b^2\le (m_H-2m_\tau)^2$) and not by  $m_H^2$,
lepton mass effects are nonnegligible in particular close to the threshold of
the off-shell decays. We calculate the rates and single angle decay
distributions and compare them with the corresponding rates and single angle
decay distributions for the nonidentical particle decays
$H\to Z^\ast(\to e^+e^-)+Z^\ast(\to\mu^+\mu^-)$ involving negligible lepton
masses.
\end{abstract}

\noindent
Keywords: decay of the Higgs boson; identical particle effects;
lepton mass effects

\newpage
\section{Introduction}
The decay channel $H\to ZZ^{(*)}\to 4\ell$ was one of the main decay channels
for the observation of the Standard Model Higgs boson in 2012 by the ATLAS and
CMS collaborations~\cite{ATLAS:2012yve,CMS:2012qbp}. Most authors who have
studied the decay $H\to\ell\ell\ell\ell$ of the Higgs boson into four leptons
(for the state of the art, see Refs.~\cite{Bredenstein:2006rh,%
Bredenstein:2006ha,Dittmaier:2012vm,Boselli:2015aha,Boselli:2017pef,%
Denner:2019fcr}) have shied away from a detailed investigation of identical
particle effects in the decay distributions of
$H\to Z^\ast Z^\ast\to\ell^+\ell^-\ell^+\ell^-$. A first appraisal of the
importance of identical particle effects~\cite{Ranft:1974yj,DeMuynck:1975sij,%
Schellekens:1980si} can be obtained from a comparison of the branching ratios
of a $125\GeV$ Higgs decaying into nonidentical and identical lepton pairs,
collected from different original works in Ref.~\cite{Denner:2011mq}. The
branching ratios listed in Ref.~\cite{Denner:2011mq} are
$B(H\to ee\mu\mu)=5.93\cdot 10^{-5}$ and $B(H\to eeee)=3.27\cdot 10^{-5}$. The
approximate factor of two between the two rates reflects (i) the statistical
factor of $1/4$ and (ii) the doubling of noninterference contributions in the
identical particle case. The small deviation of the rate ratio from the exact
value of $2$ must be assigned to the contributions of the two interference
terms. Judging from the numbers calculated in Ref.~\cite{Denner:2011mq} one
concludes that the interference contributions add constructively and are
approximately $10\,\%$ in size.   

It turns out that results for identical lepton decays of the Higgs boson,
though calculated including NLO radiative corrections, unfortunately cannot be
inferred from state-of-the-art Monte Carlo packages like
{\sc Hto4L}~\cite{Boselli:2015aha} or {\sc Prophecy4f}~\cite{Denner:2019fcr},
for two reasons. First, both packages render final state fermions massless and
use fermion masses only as regulators for mass singularities. Second, in
Ref.~\cite{Denner:2019fcr} it is stressed ``that the distributions do not
necessarily have direct physical significance if identical or invisible
particles are present in the final state.'' These two issues are the starting
points for our argumentation. In the following we point out the importance of
the lepton mass for the phase space, and we look for observables that can be
measured in experiments also for identical particle final states.

\begin{figure}
\begin{center}
\epsfig{figure=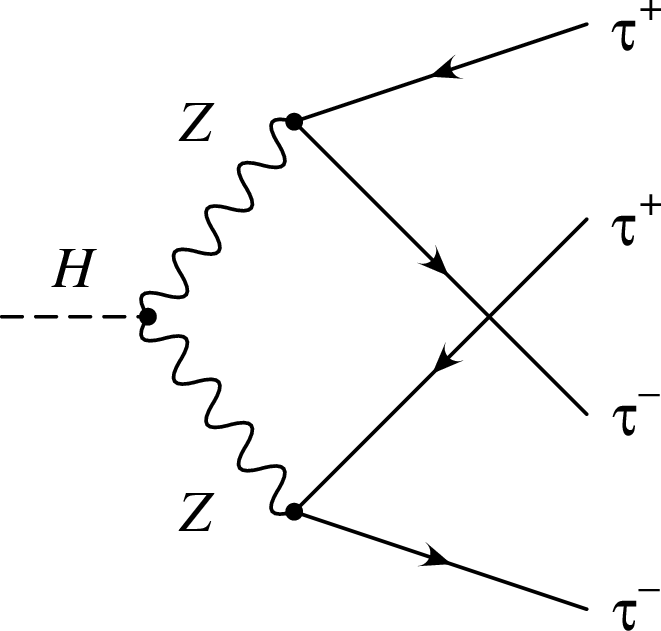, scale=0.4}\qquad
\epsfig{figure=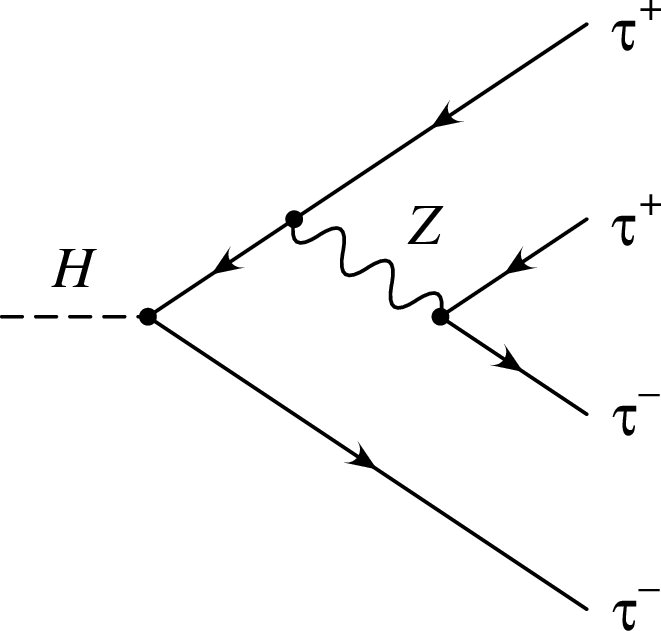, scale=0.4}\qquad
\epsfig{figure=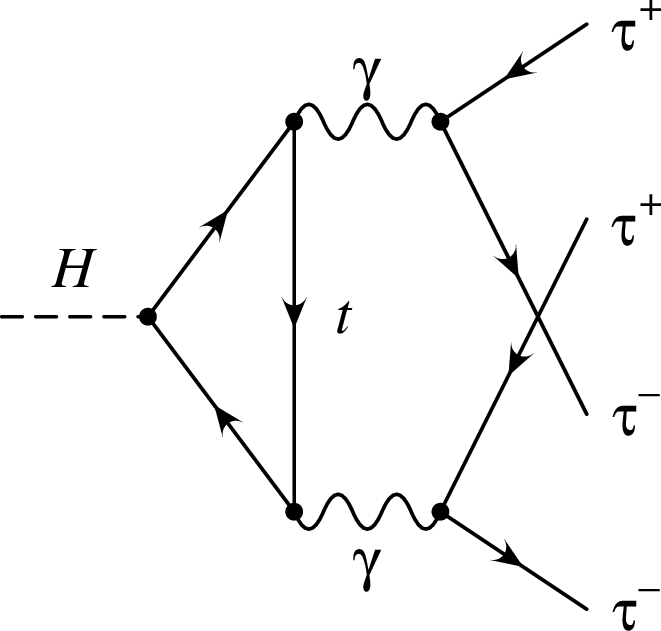, scale=0.4}\\[12pt]
(I)\kern140pt(II)\kern140pt(III)
\caption{\label{htauIII}Exemplary Feynman diagrams for the class I, II and III
  contributions to the Higgs decay process $H\to\tau^+\tau^+\tau^-\tau^-$ into
  massive leptons}
\end{center}
\end{figure}

There is a multitude of Feynman diagrams that contribute to
$H\to\ell\ell\ell\ell$ for massive leptons as for $\ell=\tau^\pm$. We divide
these into the three classes I, II and III. The first class contains the
Feynman diagrams that contribute to $H\to\ell\ell\ell\ell$ also in the zero
lepton mass case. When $m\neq 0$ one has in addition contributions
proportional to $g_{H\ell\ell}$ (class II). Finally, the class III
contributions comprise loop-induced higher order contributions such as
$H \to\gamma^\ast\gamma^\ast\to\ell\ell\ell\ell$. In Fig.~\ref{htauIII} we
present an exemplary diagram for each of the classes. Note that class III
stands also for NLO radiative corrections which are not considered in this
work in order to stay with a semi-analytical approach. These corrections are
expected to be small compared to the mass effects of $10\%$ (cf.\ Appendix~B).
The question on how to identify the $\tau$ leptons is dealt with in a honors
thesis related to CMS~\cite{Bohenick:2019}. The application of the kinematic
method presented in this thesis is out of the scope of our paper, but the
method gives hope that this identification is feasible.

The paper is organised as follows. In Sec.~2 we concentrate on the discussion
of class I contributions where we attempt to clarify several issues concerning
identical particle effects in the decay $H\to\ell\ell\ell\ell$, including
narrow width effects. In Sec.~3 we further analyse lepton mass effects in
these decays and provide numerical results for the single angle decay
distributions. In Sec.~4 we present a summary of numerical results for the
many class II contributions. In Sec.~5 we give our conclusions. The Appendices
contain details about the kinematics of the single angle decay distributions
and about class III contributions.

\section{The rate calculation
  $H\to Z^\ast(\to\tau^+\tau^-)+Z^\ast(\to\tau^+\tau^-)$}
\begin{figure}
\begin{center}
\epsfig{figure=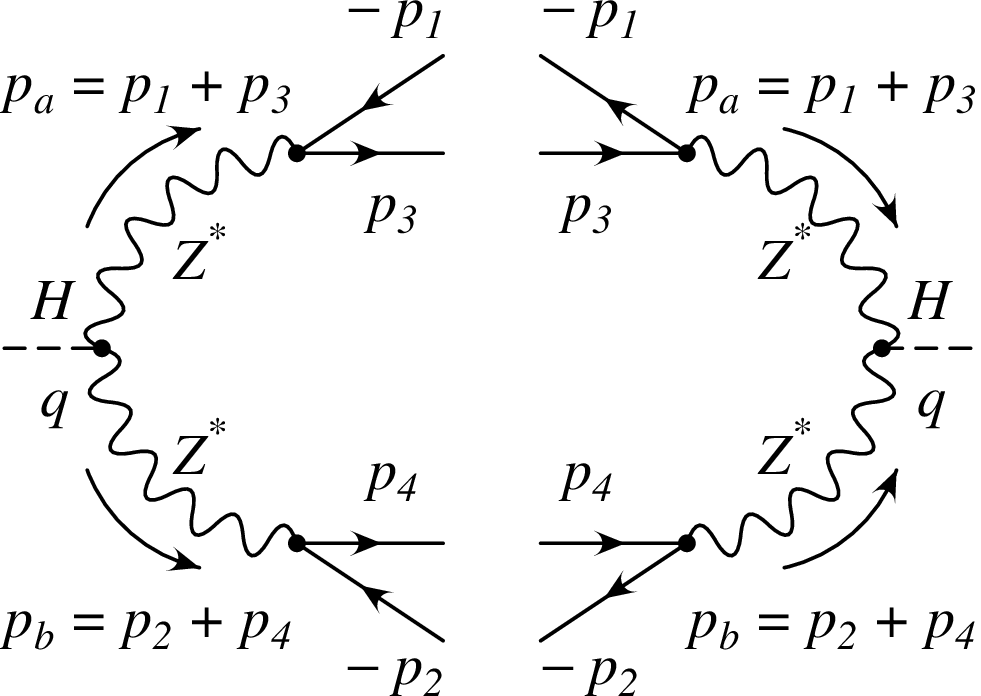, scale=0.4}\qquad
\epsfig{figure=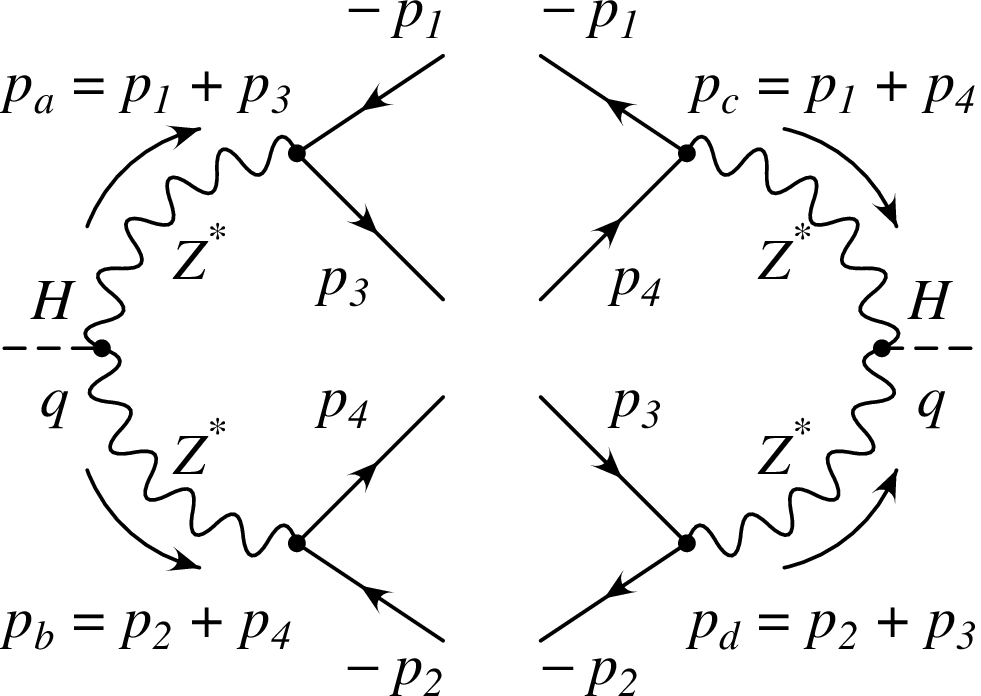, scale=0.4}\\
(a)\kern196pt(b)\\[12pt]
\epsfig{figure=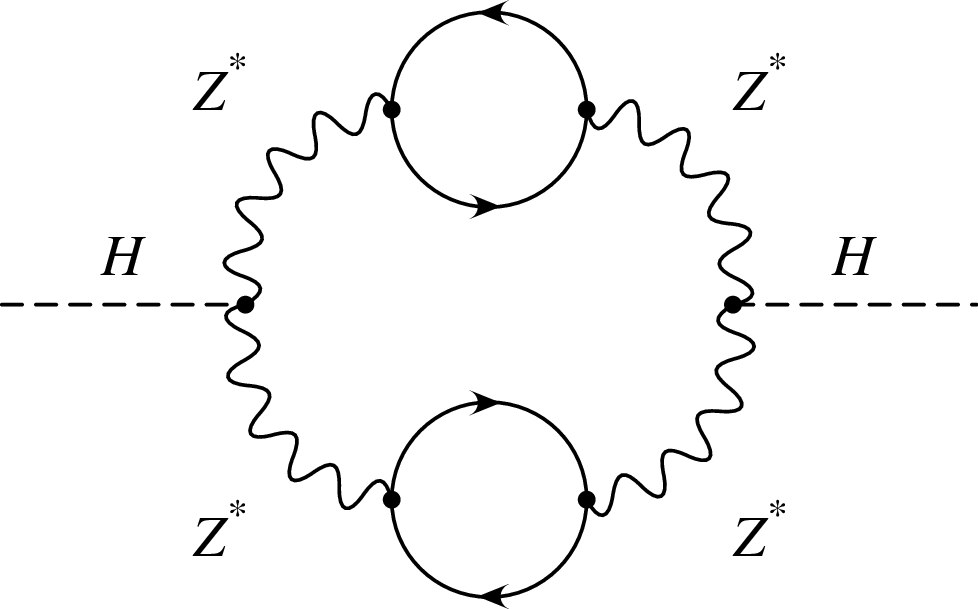, scale=0.4}\qquad
\epsfig{figure=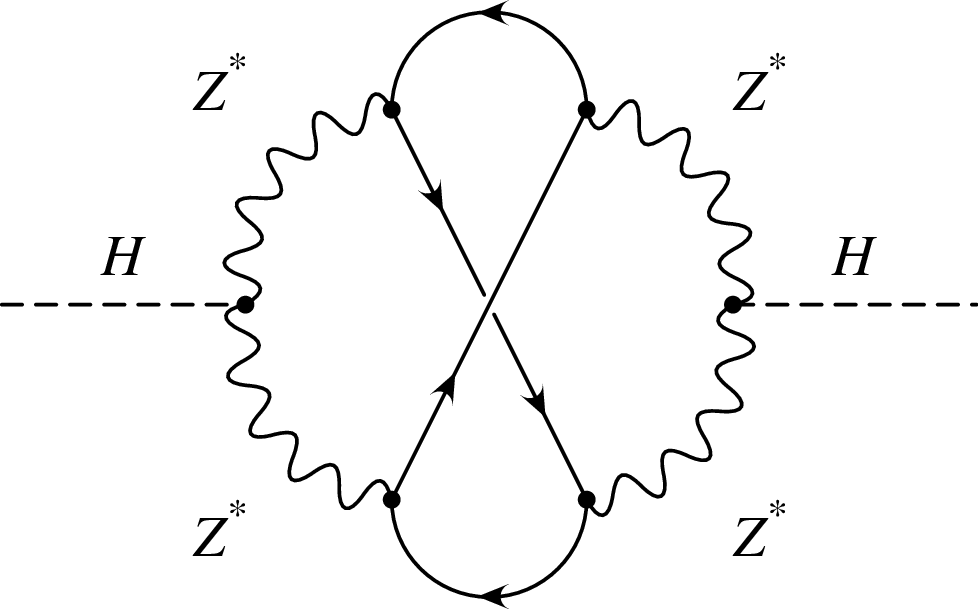, scale=0.4}\\
(c)\kern196pt(d)\\[12pt]
\caption{\label{htauab}Contributions (a) $|M_A|^2=|M_B|^2$ and (b)
$\real(M_AM_B^*)=\real(M_BM_A^*)$ from $|M_A+M_B|^2$. These contributions can
be obtained as absorptive parts of fermionic two- and one-loop diagrams via
cuts of the diagrams (c) and (d), respectively.}
\end{center}
\end{figure}

The four-body decay $H\to Z(\to\tau^+\tau^-)+Z^\ast(\to\tau^+\tau^-)$ differs
very much from e.g.\ the decay $H\to Z(\to\mu^+\mu^-)+Z^\ast(\to\tau^+\tau^-)$
in that one has to take into account interference effects resulting from the
fact that one has two pairs of identical particles in the former decay.
\begin{figure}
\begin{center}
\epsfig{figure=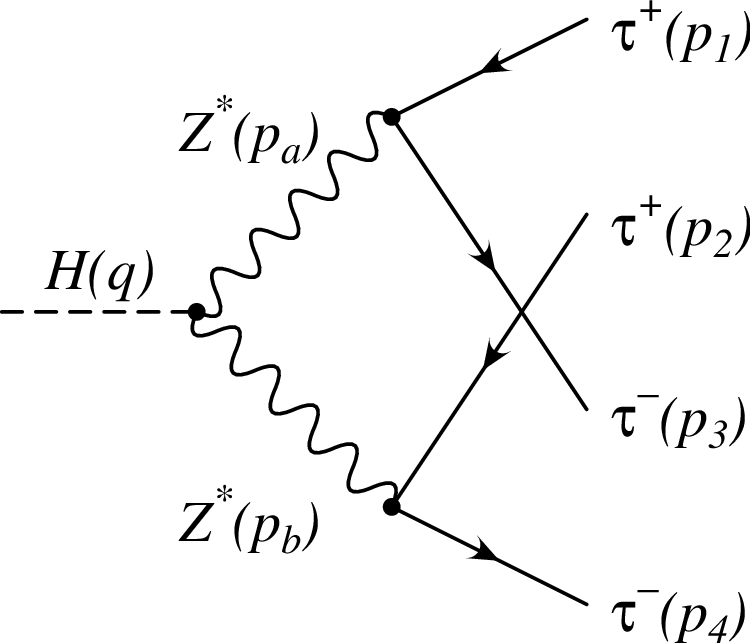, scale=0.4}\qquad\qquad
\epsfig{figure=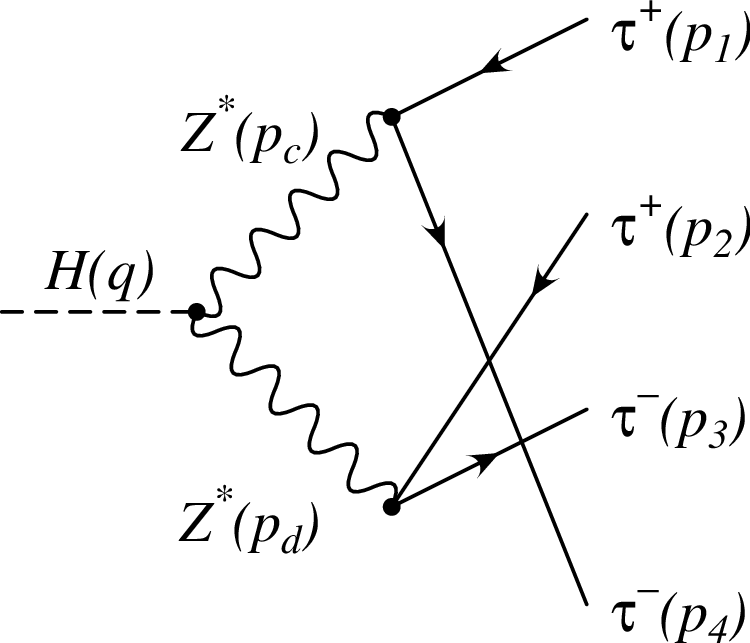, scale=0.4}\\
(A)\kern160pt(B)
\end{center}
\caption{\label{htauabp}Feynman diagrams A and B contributing to 
$H\to Z^\ast(\to\tau^+\tau^-)+Z^\ast(\to\tau^+\tau^-)$}
\end{figure}
According to the two diagrams in Fig.~\ref{htauabp} one has the two amplitudes
\begin{eqnarray}\label{AB}
M_A&=&M\left(\tau^+(p_1),\tau^-(p_3),\tau^+(p_2),\tau^-(p_4)\right)\nonumber\\
M_B&=&M\left(\tau^+(p_1),\tau^-(p_4),\tau^+(p_2),\tau^-(p_3)\right)
\end{eqnarray}
where $M_B$ is obtained from $M_A$ by exchanging the two $\tau^-$ leptons. The
two additional configurations where the $\tau^+$ leptons are exchanged or
where both $\tau^+$ and $\tau^-$ leptons are exchanged simultaneously are
topologically equivalent to the above two diagrams~(\ref{AB}) and should
therefore be discarded. When squaring the amplitudes one obtains
\begin{equation}
|M_A+M_B|^2=|M_A|^2+2\real(M_AM_B^{\ast})+|M_B|^2.
\end{equation} 
Let us add a few general remarks.
It is clear that the rate contributions of $|M_A|^2$ and $|M_B|^2$ are
identical to each other since their mutual contributions are obtained by the
exchange $p_3\leftrightarrow p_4$ which also leaves the measure of the phase
space integration invariant. In the case when one neglects the interference
contribution $2\real(M_AM_B^{\ast})$ one therefore obtains the relation
$\Gamma(H\to\tau^+\tau^-\tau^+\tau^-)=1/2\cdot\Gamma(H\to\tau^+\tau^-\mu^+\mu^-
)$ where the statistical factor $1/4$ has been taken into account in the
identical fermion case. The nondiagonal interference contribution proportional
to $2\real(M_AM_B^*)$ corresponds to the absorptive part of a fermionic
one-loop diagram compared to the fermionic two-loop diagram of the diagonal
contribution, as can be inferred from the calculation of $|M_A+M_B|^2$
illustrated in Fig.~\ref{htauab}. One must therefore be careful to include an
extra minus sign in the nondiagonal contribution. 

In the following we shall separately calculate the rate for the two 
non-interference contributions $|M_A|^2+|M_B|^2$ and the interference
contributions $2\real(M_AM_B^*)$. Henceforth we shall refer to the
non-interference contribution as the diagonal contribution and the
interference contribution as the nondiagonal contribution.

\subsection{The diagonal noninterference contribution}
The contributions of the diagonal contributions $|M_A|^2=|M_B|^2$ to 
the rate are not difficult to evaluate since they can be seen to factorize
as described in some detail in Ref.~\cite{Berge:2015jra}. Using the 
results of Ref.~\cite{Berge:2015jra} the differential rate corresponding to
the contribution of $|M_A|^2+|M_B|^2=2|M_A|^2$ can be written in the form
(including the identical particle factor of 1/4)
\begin{eqnarray}\label{offshellZZ}
\lefteqn{\frac{d\Gamma^{AA}}{dp_a^2dp_b^2}(p_a^2,p_b^2)\ =\ \frac 14\,\cdot
  \frac{2\alpha^3}{9\pi^2m_H^2}|\vec p_{Z^\ast}(p_a^2,p_b^2)| 
  \frac{|\vec p_\ell(p_a)|}{\sqrt{p_a^2}}
  \frac{|\vec p_\ell(p_b)|}{\sqrt{p_b^2}}\,
  \frac{m_Z^2}{256\sin^6\theta_W\cos^6\theta_W}}\nonumber\\&&\strut\times
  \frac1{(p_a^2-m_Z^2)^2+m_Z^2\Gamma_Z^2}\,
  \frac1{(p_b^2-m_Z^2)^2+m_Z^2\Gamma_Z^2}\,p_a^2p_b^2\nonumber\\&&\strut\times
  \Bigg\{v_\ell^2(1+2\frac{m^2}{p_a^2})P_{1\mu\nu}(p_a)
  +a_\ell^2\Big((1-4\frac{m^2}{p_a^2})P_{1\mu\nu}(p_a)
  -3a_\ell^2\cdot 2\frac{m^2}{p_a^2}F^2_S(p_a^2)P_{0\mu\nu}(p_a)
  \Big)\Bigg\}\nonumber\\&&\strut\times
  \Bigg\{v_\ell^2(1+2\frac{m^2}{p_b^2})P_1^{\mu\nu}(p_b)
  +a_\ell^2\Big((1-4\frac{m^2}{p_b^2})P_1^{\mu\nu}(p_b)
  -3a_\ell^2\cdot 2\frac{m^2}{p_b^2}F^2_S(p_b^2)P_0^{\mu\nu}(p_b)
  \Big)\Bigg\}\qquad
\end{eqnarray}
where $v_\ell=-1+4\sin^2\theta_W$ and $a_\ell=-1$ are the vector and axial
vector couplings of the $Z$ boson. $\alpha=e^2/(4\pi)$ is the fine structure
constant for which we use the value $\alpha(m_H)\approx 1/120$. For the
kinematics we have used $p_a=(E_a;\vec p_{Z^\ast})$ and
$p_b=(E_b;-\vec p_{Z^\ast})$ with
\begin{equation}
E_a=\frac{m_H^2+p_a^2-p_b^2}{2m_H},\quad
E_b=\frac{m_H^2-p_a^2+p_b^2}{2m_H},\quad
|\vec p_{Z^\ast}|=\frac1{2m_H}\sqrt{\lambda(m_H^2,p_a^2,p_b^2)},
\end{equation}
and $p_ap_b=\frac12(m_H^2-p_a^2-p_b^2)$, where
$\lambda(a,b,c):=a^2+b^2+c^2-2ab-2ac-2bc$ is the K\"all\'en function. The
remaining phase space factors $|\vec p_1|=|\vec p_3|=|\vec p_\ell(p_a)|$ and
$|\vec p_2|=|\vec p_4|=|\vec p_\ell(p_b)|$ are calculated in the respective
rest frames of the decaying vector bosons and read
\begin{equation}
|\vec p_\ell(p_a)|=\frac12\sqrt{p_a^2-4m^2}=:\frac12\sqrt{p_a^2}\,v_a,
  \qquad
|\vec p_\ell(p_b)|=\frac12\sqrt{p_b^2-4m^2}=:\frac12\sqrt{p_b^2}\,v_b.
\end{equation}

Note the appearance of spin-1 and spin-0 projectors given by 
\begin{equation}\label{offshell}
P_1^{\mu\nu}(p_a)=g^{\mu\nu}-\frac{p_a^\mu p_a^\nu}{p_a^2},\qquad
P_0^{\mu\nu}(p_a)=\frac{p_a^\mu p_a^\nu}{p_a^2}.
\end{equation}
and similarly $P_1^{\mu\nu}(p_b)$ and $P_0^{\mu\nu}(p_b)$. The factor
$F_S(p^2)=1-p^2/m_Z^2$ multiplying the scalar contribution in
Eq.~(\ref{offshellZZ}) is a result of having used the unitary gauge for the
gauge boson propagator (see Ref.~\cite{Berge:2015jra}). The spin-0 piece of
the unitary gauge boson propagators carries the helicity flip factors
$m^2/p_a^2$ and $m^2/p_b^2$ which we need to take into account when
discussing lepton mass effects in the decay $H\to\ell\ell\ell\ell$. The
relevant Lorentz contractions in Eq.~(\ref{offshellZZ}) can be calculated to be 
\begin{eqnarray}
P_1^{\mu\nu}(p_a)P_{1\,\mu\nu}(p_b)
  &:=&\rho_{TT}(p_a^2,p_b^2)\ =\ 2+\frac{(p_ap_b)^2}{p_a^2p_b^2}
  \ =\ 3+\frac{m_H^2|\vec p_{Z^\ast}|^2}{p_a^2p_b^2},\label{contract11}\\
P_1^{\mu\nu}(p_a)P_{0\,\mu\nu}(p_b)
  &:=&\rho_{TS}(p_a^2,p_b^2)\ =\ 1-\frac{(p_ap_b)^2}{p_a^2p_b^2}\ 
  \ =\ -\frac{m_H^2|\vec p_{Z^\ast}|^2}{p_a^2p_b^2},\qquad\label{contract10}\\
P_0^{\mu\nu}(p_a)P_{0\,\mu\nu}(p_b)
  &:=&\rho_{SS}(p_a^2,p_b^2)\ =\ \frac{(p_ap_b)^2}{p_a^2p_b^2}
  \ =\ 1+\frac{m_H^2|\vec p_{Z^\ast}|^2}{p_a^2p_b^2}.\label{contract00} 
\end{eqnarray}
Following the standard convention we label the components of the $1\otimes1$
spin--spin density matrix elements in Eq.~(\ref{contract11}) by labels $T$
(transverse) and $S$ (scalar). The scalar--scalar contribution $\rho_{SS}$ in
Eq.~(\ref{offshellZZ}) appears multiplied with the product
$m^2/p_a^2\cdot m^2/p_b^2$ of helicity flip factors and can be neglected for
all practical purposes. 

The contractions~(\ref{contract11}), (\ref{contract10}) and~(\ref{contract00})
stand for spin--spin density matrix elements which determine the angular
coefficients of the angular decay distributions of the subsequent decays
$Z^\ast\to\ell\ell$~\cite{Berge:2015jra}. The contractions have been written
in two different forms. The first equations are suitable for a discussion of
the large recoil region where $p_a^2$ and $p_b^2$ are small. The second
equations are suitable for the low recoil region where $|\vec p_{Z^\ast}|$ is
small. In the large recoil region the dominant contributions can be seen to be
given by $\rho_{TT}\approx\rho_{TS}\approx\rho_{SS}$, whereas on has
$\rho_{SS}\approx 3\rho_{TT}$ and $\rho_{TS}\approx 0$ in the low recoil
region. 

The rate is obtained from Eq.~(\ref{offshellZZ}) by $p_a^2$ and $p_b^2$
integration according to
\begin{equation}\label{rate1}
\Gamma^{AA}\ =\ \int_{4m^2}^{(m_H-2m)^2}dp_b^2
  \int_{4m^2}^{(m_H-\sqrt{p_b^2})^2}dp_a^2\quad
  \frac{d\Gamma^{AA}}{dp_a^2dp_b^2}(p_a^2,p_b^2).
\end{equation}
The integrations in Eq.~(\ref{rate1}) can be performed numerically by using
MATHEMATICA. The results are given in Table~\ref{tabrate} where we list the 
integrated rates for the three cases $\ell=e,\mu,\tau$. For the mass of the
Higgs boson we use the central value $m_H=125.09\pm0.24\GeV$~\cite{Aad:2015zhl}.

\subsection{The nondiagonal interference contribution}
The rate calculation corresponding to the interference contribution 
$\real(M_AM_B^{\ast})$ is considerably more difficult. One reason is that the
integrand does not factorize into $p_a$- and $p_b$-side contributions as in 
the diagonal case (see Eq.~(\ref{offshellZZ})). As a result the requisite 
angular integrations can no longer be done analytically as was possible in 
the diagonal case. 

\begin{figure}
\begin{center}
\epsfig{figure=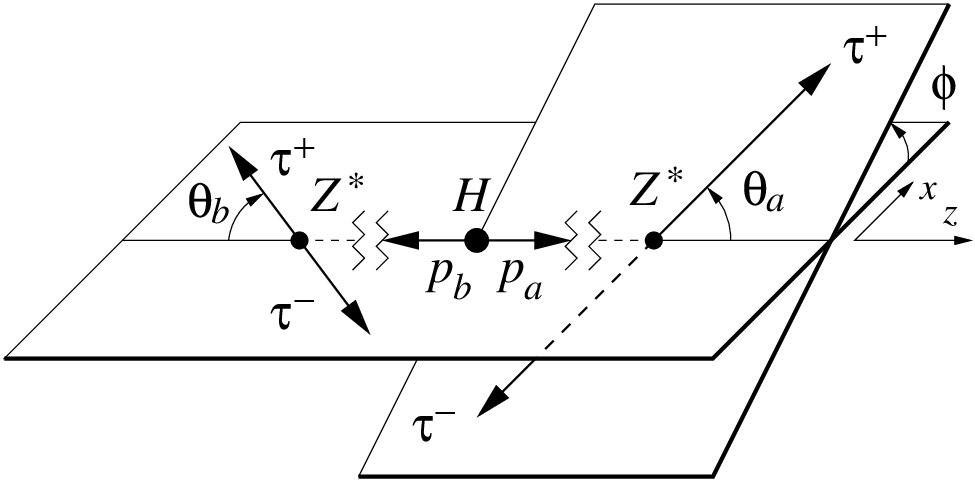, scale=0.7}
\caption{\label{planesab}Definition of the momenta $p_a$ and $p_b$, the polar
angles $\theta_a$ and $\theta_b$, and the azimuthal angle $\phi$ in the
cascade decay $H\to Z^\ast(\to\tau^+\tau^-)+Z^\ast(\to\tau^+\tau^-)$}
\end{center}
\end{figure}

We begin by setting up the five-dimensional phase space. As in
Ref.~\cite{Cabibbo:1965zzb,Pais:1968zza} (see also
Ref.~\cite{Cappiello:2011qc}) we choose the following five phase space
variables: the two invariant masses $p_a^2=(p_1+p_3)^2$ and
$p_b^2=(p_2+p_4)^2$, the two polar angles $\theta_a$ and $\theta_b$, and the
azimuthal angle $\phi$ (also called acoplanarity, see Fig.~\ref{planesab}).
For the twice differential rate corresponding to the nondiagonal interference
contribution $2\real(M_AM_B^{\ast})$ one obtains (including the identical
particle factor of 1/4)
\begin{eqnarray}\label{nondiag}
\lefteqn{\frac{d\Gamma^{AB}}{dp_a^2dp_b^2}(p_a^2,p_b^2)\ =\ \frac 14\,\cdot
  \frac{\alpha^3}{(2\pi)^3m_H^2}|\vec p_{Z^\ast}(p_a^2,p_b^2)| 
  \frac{|\vec p_\ell(p_a)|}{\sqrt{p_a^2}}
  \frac{|\vec p_\ell(p_b)|}{\sqrt{p_b^2}}\,
  \frac{m_Z^2}{256\sin^6\theta_W\cos^6\theta_W}}\\&&\strut\times
  \int d\cos\theta_ad\cos\theta_bd\phi
  \Big(v_\ell^4N_0+v_\ell^2a_\ell^2N_2+a_\ell^4N_4\Big)\nonumber\\  
&&\strut\times
  \frac{D_aD_bD_cD_d+m_Z^2\Gamma_Z^2
  (-D_aD_b+D_aD_c+D_aD_d+D_bD_c+D_bD_d
  -D_cD_d)+m_Z^4\Gamma_Z^4}{(D_a^2+m_Z^2\Gamma_Z^2)
  (D_b^2+m_Z^2\Gamma_Z^2)(D_c^2+m_Z^2\Gamma_Z^2)(D_d^2+m_Z^2\Gamma_Z^2)}\,,
\nonumber
\end{eqnarray}
where the numerator factors $N_0,N_2$ and $N_4$ are too long to be 
presented here.

The pole factors $D_i$ $i=a,b,c,d$ in Eq.~(\ref{nondiag}) read 
\begin{eqnarray}
D_a&=&(p_1+p_3)^2-m_Z^2\ =\ p_a^2-m_Z^2,\nonumber\\[7pt]
D_b&=&(p_2+p_4)^2-m_Z^2\ =\ p_b^2-m_Z^2,\nonumber\\[7pt]
D_c&=&(p_1+p_4)^2-m_Z^2\ =\ p_c^2-m_Z^2\ =\ 2m^2-m_Z^2\nonumber\\[3pt]&&
  +\frac14\Big((m_H^2-p_a^2-p_b^2)(1-\cos\theta_a\cos\theta_bv_av_b)
  -2\sqrt{p_a^2p_b^2}\cos\phi\sin\theta_a\sin\theta_bv_av_b
\nonumber\\&&\strut
  +\sqrt{\lambda(m_H^2,p_a^2,p_b^2)}(\cos\theta_av_a-\cos\theta_bv_b)\Big),
  \nonumber\\[3pt]
D_d&=&(p_2+p_3)^2-m_Z^2\ =\ p_d^2-m_Z^2\ =\ 2m^2-m_Z^2\nonumber\\[3pt]&&
  +\frac14\Big((m_H^2-p_a^2-p_b^2)(1-\cos\theta_a\cos\theta_bv_av_b)
  -2\sqrt{p_a^2p_b^2}\cos\phi\sin\theta_a\sin\theta_bv_av_b\nonumber\\&&\strut
  -\sqrt{\lambda(m_H^2,p_a^2,p_b^2)}(\cos\theta_av_a-\cos\theta_bv_b)\Big),
\end{eqnarray}
where $D_c$ and $D_d$ depend on the angles which is the second reason for not
being able to obtain an analytical result. The cosines of the angles are given
by
\begin{equation}
\cos\theta_a=\frac{\vec p_a\cdot\vec p_1}{|\vec p_a||\vec p_1|},\quad
\cos\theta_b=\frac{\vec p_b\cdot\vec p_2}{|\vec p_b||\vec p_2|},\quad
\cos\phi=\frac{\vec p_1^\perp\cdot\vec p_2^\perp}{|\vec p_1^\perp|
  |\vec p_2^\perp|}
\end{equation}
where $\vec p_i^\perp$ is the component of $\vec p_i$ perpendicular to
$\vec p_a$ (or $\vec p_b$, respectively), given by
\begin{equation}
\vec p_1^\perp=\vec p_1-\frac{(\vec p_1\cdot\vec p_a)\vec p_a}{|\vec p_a|^2},
\qquad
\vec p_2^\perp=\vec p_2-\frac{(\vec p_2\cdot\vec p_b)\vec p_b}{|\vec p_b|^2}.
\end{equation}
in the rest frame of the decaying $Z$ bosons.

Note that due to the crossing of momenta, for the nondiagonal interference
contribution it is necessary to switch to a more general notation involving a
set of four invariant masses $p_a^2$, $p_b^2$, $p_c^2$ and $p_d^2$ instead of
the initial first two of this set on which they depend via the angles. The
input for this calculation and the kinematics necessary for it is transferred
to the numerical integration routine VEGAS~\cite{Lepage:2020tgj}. The
kinematics is expressed in terms of four-vectors in the rest frames of the
decaying $Z$ bosons with polar angles $\theta_a$ and $\theta_b$, boosted to
the rest frame of the Higgs boson via the rapidities
\begin{equation}
\lambda_a=\arctanh\pfrac{\sqrt{\lambda(m_H^2,p_a^2,p_b^2)}}{m_H^2+p_a^2-p_b^2},
\qquad
\lambda_b=-\arctanh\pfrac{\sqrt{\lambda(m_H^2,p_a^2,p_b^2)}}{m_H^2-p_a^2+p_b^2},
\end{equation}
and turned around the $z$ axis through $\phi$, resulting in
\begin{eqnarray}
p_{1/3}&=&\frac12\sqrt{p_a^2}
  \Big(\cosh\lambda_a\pm v_a\cos\theta_a\sinh\lambda_a;
  \pm v_a\sin\theta_a\cos\phi,\pm v_a\sin\theta_a\sin\phi,
  \nonumber\\&&\qquad\qquad
  \sinh\lambda_a\pm v_a\cos\theta_a\cosh\lambda_a\Big),\nonumber\\
p_{2/4}&=&\frac12\sqrt{p_b^2}
  \left(\cosh\lambda_b\pm v_b\cos\theta_b\sinh\lambda_b;\pm v_b\sin\theta_b,
  0,\sinh\lambda_b\pm v_b\cos\theta_b\cosh\lambda_b\right).\qquad
\end{eqnarray}

\subsection{The phase space for $p_a^2$ and $p_b^2$}
\begin{figure}\begin{center}
\epsfig{figure=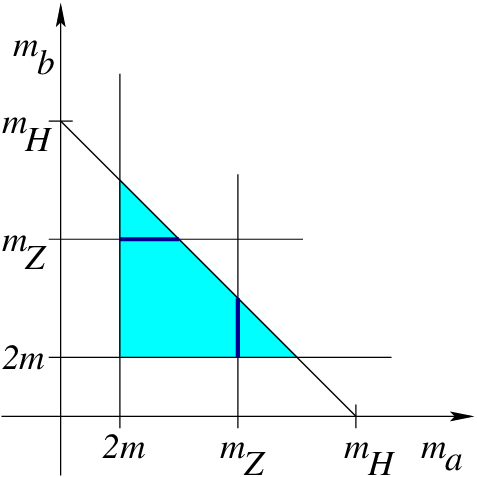, scale=0.8}
\caption{\label{phasesph}Phase space domain for the invariant masses $m_a$ and
$m_b$ in dependence on the Higgs boson mass $m_H$, as restricted by the three
straight lines $m_a=2m$, $m_b=2m$ and $m_a+m_b=m_H$ and painted in light blue.
The localizations of the two $Z$ poles at $m_a=m_Z$ and $m_b=m_Z$ are
indicated by two line segments in dark blue cutting the phase space domain.
Note that the diagram is used for illustrative reasons only and not fixed to
the physical masses $m=m_\tau$, $m_Z$ and $m_H$.}
\end{center}\end{figure}

The phase space domain in $(p_a^2,p_b^2)$ to be integrated over can be
investigated by looking at the phase space. As pointed out in
Ref.~\cite{Kaldamae:2014fua}, the boundary of the phase space domain can be
found by demanding that the measure is real, i.e.\ all the radicands are
positive. Claiming that
\begin{equation}
1-\frac{4m^2}{p_a^2}\ge 0,\qquad
1-\frac{4m^2}{p_b^2}\ge 0,\qquad
\lambda(m_H^2,p_a^2,p_b^2)\ge 0,
\end{equation}
the first two inequalities can be resolved to $p_a^2,p_b^2\ge 4m^2$. For
the last one we replace $p_a^2$ and $p_b^2$ by the squared invariant masses
$m_a^2$ and $m_b^2$ and obtain
\begin{equation}
(m_H^2-(m_a+m_b)^2)(m_H^2-(m_a-m_b)^2)>0
\end{equation}
The only ``physical'' restriction is given by $m_a+m_b\le m_H$ which is also
very intuitive. Therefore, the phase space domain is given by the intersection
of $m_a\ge 2m$, $m_b\ge 2m$ and $m_a+m_b\le m_H$ as shown in
Fig.~\ref{phasesph}.
In terms of $p_a^2$ and $p_b^2$ the integrations limits are given by
\begin{equation}
4m^2\le p_a^2\le(m_H-2m)^2,\qquad
4m^2\le p_b^2\le(m_H-\sqrt{p_a^2})^2.
\end{equation}
In case of the narrow width approximation (NWA), the phase space domain is
restricted to the two line segments along $p_a^2=m_Z^2$ and $p_b^2=m_Z^2$ and
indicated in Fig.~\ref{phasesph} in dark blue, cutting the light blue phase
space domain. If for instance the first $Z$ boson is on shell, $m_a^2=m_Z^2$,
the phase space is the vertical line segment limited by
$4m^2\le p_b^2\le(m_H-m_Z)^2$.

As suggested by Je\.zabek and K\"uhn~\cite{Jezabek:1988iv,Jezabek:1993wk},
singularities at $p_a^2=m_Z^2$ and $p_b^2=m_Z^2$ (and eventually singularities
occuring in connection with $p_c^2$ and $p_d^2$) are defended by adding a
Breit--Wigner contribution to the denominator factor, for
instance\footnote{A more recent treatment of this topic can be found e.g.\ in
Ref.~\cite{Denner:2019vbn}.}
\begin{equation}
D_a:=(p_a^2-m_Z^2)\to(p_a^2-m_Z^2)+im_Z\Gamma_Z=:D_a+i\eps.
\end{equation}
One obtains
\begin{equation}
|M_A|^2=\frac{N^{AA}}{(D_a+i\eps)(D_b+i\eps)(D_a-i\eps)(D_b-i\eps)}
  =\frac{N^{AA}}{(D_a^2+\eps^2)(D_b^2+\eps^2)}.
\end{equation}
The nondiagonal interference contribution is more complicated. Taking the
numerator to be symbolically $N^{AB}=N_r^{AB}+iN_i^{AB}$, the product
\begin{equation}
M_AM_B^*=\frac{N_r^{AB}+iN_i^{AB}}{(D_a+i\eps)(D_b+i\eps)(D_c-i\eps)
  (D_d-i\eps)},
\end{equation}
leads to
\begin{equation}
\real(M_AM_B^*)=\frac{\real\left((D_a-i\eps)(D_b-i\eps)(D_c+i\eps)(D_d+i\eps)
  (N_r^{AB}+iN_i^{AB})\right)}{(D_a^2+\eps^2)(D_b^2+\eps^2)(D_c^2+\eps^2)
  (D_d^2+\eps^2)}.
\end{equation}
The numerical results for the nondiagonal contribution $\Gamma^{AB}$ obtained
by VEGAS is shown in Table~\ref{tabrate} for the three leptons $\ell=e,\mu,\tau$.

\begin{table}[ht]\begin{center}
\caption{\label{tabrate}Diagonal and nondiagonal contributions $\Gamma^{AA}$
and $\Gamma^{AB}$ in units of $10^{-7}\GeV$ to the decay
$H\to Z^\ast(\to\ell^+\ell^-)+Z^\ast(\to\ell^+\ell^-)$ for the three leptons
$\ell=e,\mu,\tau$}
\vspace{12pt}
\begin{tabular}{|r||c|c|}\hline
&$\Gamma^{AA}$&$\Gamma^{AB}$\\\hline
$\ell=e$&$2.4195(2)$&$0.24835(2)$\\
$\ell=\mu$&$2.4191(2)$&$0.24833(2)$\\
$\ell=\tau$&$2.3316(2)$&$0.24316(2)$\\\hline
\end{tabular}
\end{center}\end{table}

\subsection{Narrow width approximation}
We have mentioned before that the nondiagonal interference contribution
$\Gamma^{AB}\sim\real(M_AM_B^*)\kern-6pt$ to the rate
$\Gamma=\Gamma^{AA}+\Gamma^{AB}$ is suppressed relative to the diagonal
non-interference contribution $\Gamma^{AA}\sim|M_A|^2$. Technically this comes
about by the fact that there is a phase space momentum mismatch between the
peaking regions of diagram A and diagram B. This mismatch becomes larger as
the width $\Gamma_Z$ becomes smaller. The net result is that the phase space
integration of the nondiagonal contribution tends to a constant value
independent of the width $\Gamma_Z$. This is illustrated in
Table~\ref{tabrateratio} where we list the numerical values of the rate ratios
of the nondiagonal and diagonal contributions for different values of the $Z$
width.

\begin{table}[t]
\caption{\label{tabrateratio}Dependence of the rate ratios of diagonal and
nondiagonal contributions $\Gamma^{AA}$ and $\Gamma^{AB}$ on the $Z$ width for
the decay $H\to Z^\ast(\to\tau^+\tau^-)+Z^\ast(\to\tau^+\tau^-)$. The fourth
column gives the rate ratio in absolute values and the fifth column in units
of $\Gamma_Z/m_Z$.}
\begin{center}\begin{tabular}{lcccc}\hline 
$\Gamma_Z$&$\Gamma^{AA}$&$\Gamma^{AB}$&
$\Gamma^{AB}/\Gamma^{AA}$&
$\Gamma^{AB}/\Gamma^{AA}$
 \\
$[\rm{GeV}]$&[\rm{GeV}]&[\rm{GeV}]&&$[\Gamma_Z/m_Z]$
 \\ \hline 
$2.4952$&$2.3316(2)\cdot 10^{-7}$&$2.4316(2)\cdot 10^{-8}$ 
  &$10.4\,\%$&$3.81$\\
$1.0$&$5.1974(5)\cdot 10^{-7}$&$2.4938(2)\cdot 10^{-8}$
  &$4.8\,\%$&$4.38$\\
$0.5$&$9.958(1)\cdot 10^{-7}$&$2.5145(3)\cdot 10^{-8}$
  &$2.5\,\%$&$4.61$\\
$0.2$&$2.4226(3)\cdot 10^{-6}$&$2.5266(3)\cdot 10^{-8}$
  &$1.04\,\%$&$4.76$\\
$0.1$&$4.795(1)\cdot 10^{-6}$&$2.5303(3)\cdot 10^{-8}$
  &$0.53\,\%$&$4.81$\\
$0.05$&$9.510(3)\cdot 10^{-6}$&$2.5321(4)\cdot 10^{-8}$
  &$0.27\,\%$&$4.86$\\
\hline
\end{tabular}\end{center}
\end{table}

The entries of Table~\ref{tabrateratio} indicate that the width dependence of
the rate ratio tends to $\Gamma^{AB}/\Gamma^{AA}\sim\Gamma_Z/m_Z$ as
$\Gamma_Z \to 0$. The limiting behavior of the $\Gamma_Z$ dependence of this
rate ratio can in fact be analyzed quantitatively with the help of the
$\delta$ distribution representation
\begin{equation}
\lim_{\Gamma_Z \to 0}\,\frac1{(p_a^2-m_Z^2)^2+m_Z^2\Gamma_Z^2}
  =\frac\pi{m_Z\Gamma_Z}\,\delta(p_a^2-m_Z^2).
\end{equation}
In order to extract the narrow width dependence of the rate it is sufficient
to analyze the peaking region close to e.g.\ $p_a^2=m_Z^2$. We first discuss
the diagonal contribution to the rate. Close to $p_a^2=m_Z^2$ the integral
corresponding to the diagonal contribution can be cast into the form
\begin{eqnarray}\label{limdiag}
\lefteqn{\lim_{\Gamma_Z\to 0}\int dp_a^2\frac{F(p_a^2)}{(p_a^2-m_Z^2)^2
  +m_Z^2\Gamma_Z^2}\ =\ \frac1{m_Z\Gamma_Z}\quad\lim_{\Gamma_Z\to 0}\int dp_a^2
  \frac{m_Z\Gamma_Z\,F(p_a^2)}{(p_a^2-m_Z^2)^2+m_Z^2\Gamma_Z^2}}\nonumber\\
  &=&\frac\pi{m_Z\Gamma_Z}\,\int dp_a^2\,\delta(p_a^2-m_Z^2)\,F(p_a^2)
  \ =\ \frac{\pi}{m_Z\Gamma_Z}\,\,F(m_Z^2),\qquad\qquad
\end{eqnarray}
where the function $F(p_a^2)$ is regular at $p_a^2=m_Z^2$.

The nondiagonal contribution corresponding to Eq.~(\ref{nondiag}) is less 
singular for $p_a^2\to m_Z^2$. One only has a single Breit--Wigner pole
denominator that determines the functional behavior close to $p_a^2=m_Z^2$
which we write as
\begin{equation}
\frac1{(p_a^2-m_Z^2)-im_Z\Gamma_Z}
  =\frac{(p_a^2-m_Z^2)+im_Z\Gamma_Z}{(p_a^2-m_Z^2)^2+m_Z^2\Gamma_Z^2}
\end{equation}
for the contribution of e.g.\ $M_AM_B^*$. The imaginary part is cancelled by
the conjugate contribution $M_A^*M_B$. The structure of the $p_a^2$ integral
is now given by
\begin{equation}\label{limnondiag}
\lim_{\Gamma_Z\to 0}\,\int dp_a^2\frac{(p_a^2-m_Z^2)F_1(p_a^2)}{(p_a^2-m_Z^2)^2}
  =\frac\pi{m_Z\Gamma_Z} \,\int dp_a^2\delta(p^a_a-m_Z^2)(p_a^2-m_Z^2)
  F_1(p_a^2)=0.
\end{equation}
The phase space integration over the single pole region $p_a^2\sim m_Z^2$
thus tends to zero as $\Gamma_Z \to 0$. The integration over the remaining
phase space region results in a constant value independent of the $Z$ boson
width.

The upshot of our analysis is that one has
$\lim_{\Gamma_Z\to 0}\Gamma^{AA}\sim m_Z/\Gamma_Z$
for the diagonal contribution whereas the nondiagonal contribution tends
to a constant value, i.e.\ $\Gamma^{AB}/\Gamma^{AA}\sim\Gamma_Z/ m_Z$ as also
indicated in Table~\ref{tabrateratio}. The limiting proportionality factor
linking the two quantities can be read off Table~\ref{tabrateratio} and is
given by $\Gamma^{AB}/\Gamma^{AA}\approx 2.74\,\Gamma_Z/m_Z$.

The investigation in this subsection was prompted by an misleading statement
in the literature that the nondiagonal interference contribution is suppressed
at $O(\Gamma_Z^2/m_Z^2)$~\cite{Kniehl:1990yb}.

\section{Single angle decay distributions}
With two different pairs of particles discernible in experiments as in e.g.\
Ref.~\cite{Berge:2015jra} we would be able to distinguish between the two
channels, reconstruct the momenta $p_a$ and $p_b$, and measure the relative
angles $\theta_a$ and $\theta_b$. However, with two pairs of identical
particles this is not possible any more. A distinction can of course be made
between positively and negatively charged leptons. If the $Z$ bosons are
visible in the experiment, the angles $\theta^a:=\theta_{13}$,
$\theta^b:=\theta_{24}$, $\theta^c:=\theta_{14}$ and $\theta^d:=\theta_{23}$
between all couples of differently charged leptons can be measured. In
addition, the momenta $p_{ij}=p_i+p_j$ for the partial channels of these
couples can be reconstructed, the ladder being the momenta $p_a$, $p_b$, $p_c$
and $p_d$ already introduced before. In this case, one can measure
differential decay rates $d\Gamma/dp_i^2d\cos\theta^i$ ($i=a,b,c,d$). However,
not knowing the actual $Z$ boson channel, the observable accessible in
experiment is the mean value
\begin{equation}\label{Gamabcd}
  \frac{d\Gamma}{dp^2d\cos\theta}=\frac14\left(
  \frac{d\Gamma}{dp_a^2d\cos\theta^a}+\frac{d\Gamma}{dp_b^2d\cos\theta^b}
  +\frac{d\Gamma}{dp_c^2d\cos\theta^c}+\frac{d\Gamma}{dp_d^2d\cos\theta^d}
  \right).
\end{equation}
$p^2$ stands for the measured momentum square of the reconstructed $Z$ while
$\theta$ is the opening angle of the $\tau$ pair for which the $Z$ boson is
reconstructed. The four separate parts in Eq.~(\ref{Gamabcd}) are referred to
as $a$, $b$, $c$ and $d$ channel contributions. By using
\begin{equation}
\frac{df}{dy}\Big|_{y=g(x)}=\int\frac{df}{dy}\delta\left(y-g(x)\right)dy
  =\int\frac{df}{dx}\delta\left(g(x)-y\right)dx,
\end{equation}
the differential decay rates can be calculated to obtain
\begin{eqnarray}\label{dGami}
\frac{d\Gamma}{dp_i^2d\cos\theta^i}
  &=&\int\frac{d\Gamma}{dp_a^2dp_b^2d\cos\theta_ad\cos\theta_bd\phi}
  \delta\left(p_i^2(p_a^2,p_b^2,\theta_a,\theta_b,\phi)-p_i^2\right)
  \strut\nonumber\\&&\strut\times
  \delta\left(\cos\theta^i(p_a^2,p_b^2,\theta_a,\theta_b,\phi)
  -\cos\theta^i\right)dp_a^2dp_b^2d\cos\theta_ad\cos\theta_bd\phi,\qquad
\end{eqnarray}
where $p_i^2(p_a^2,p_b^2,\theta_a,\theta_b,\phi)$ and
$\cos\theta^i(p_a^2,p_b^2,\theta_a,\theta_b,\phi)$ are the explicit
expressions for the squa\-red channel momenta and cosines in terms of the
kinematic quantities $p_a^2$, $p_b^2$, $\theta_a$, $\theta_b$ and $\phi$.
The kinematical details are left to Appendix~A. In Fig.~\ref{mdpsxxx} the
results of the VEGAS integration is shown for different squares energies
$p^2=50$, $100$, $200$ and $500\GeV^2$. For small $p^2$ the function is peaked
close to the upper limit $\cos\theta=+1$. A detailed analysis shows that the
first peak at lower values of $\cos\theta$ is proliferated by the diagonal
noninterference contribution while the second peak close to the threshold
already present in the diagonal noniterference contribution is enhanced by the
nondiagonal interference contribution.

\begin{figure}
\begin{center}
\epsfig{figure=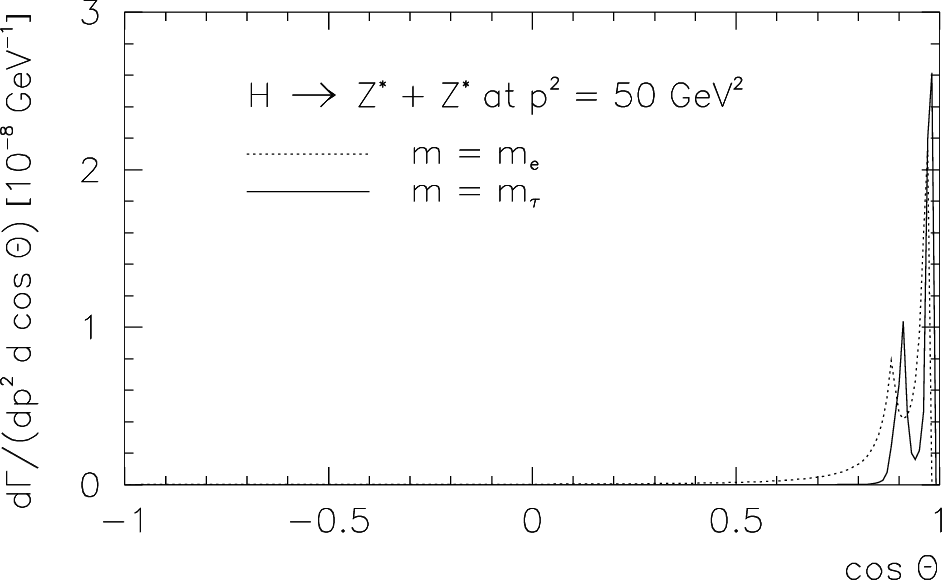,scale=0.8}
\vspace{12pt}
\epsfig{figure=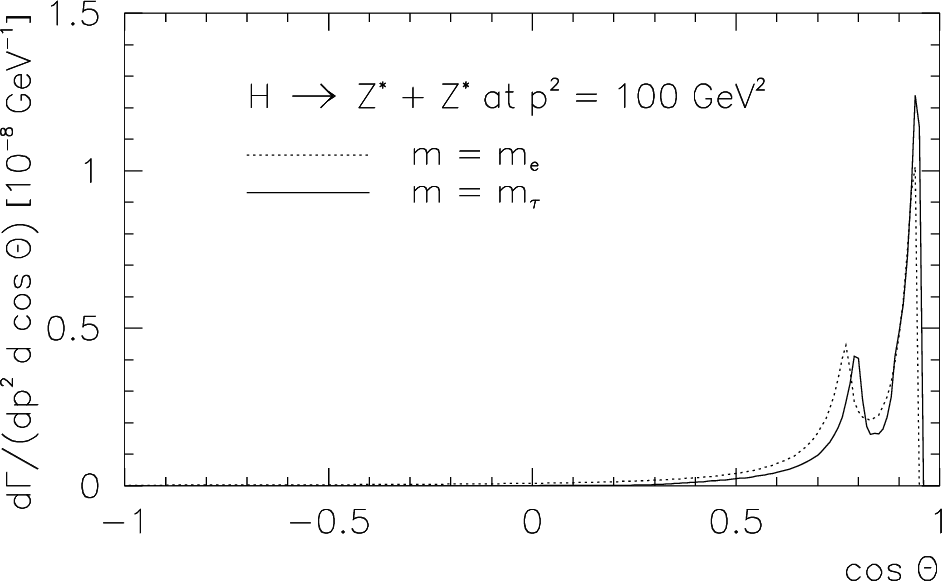,scale=0.8}
\end{center}
\caption{\label{mdpsxxx}Angular dependence of $d\Gamma/(dp^2d\cos\theta)$
for different squared energies $p^2$}
\end{figure}

\begin{figure}\addtocounter{figure}{-1}
\begin{center}
\epsfig{figure=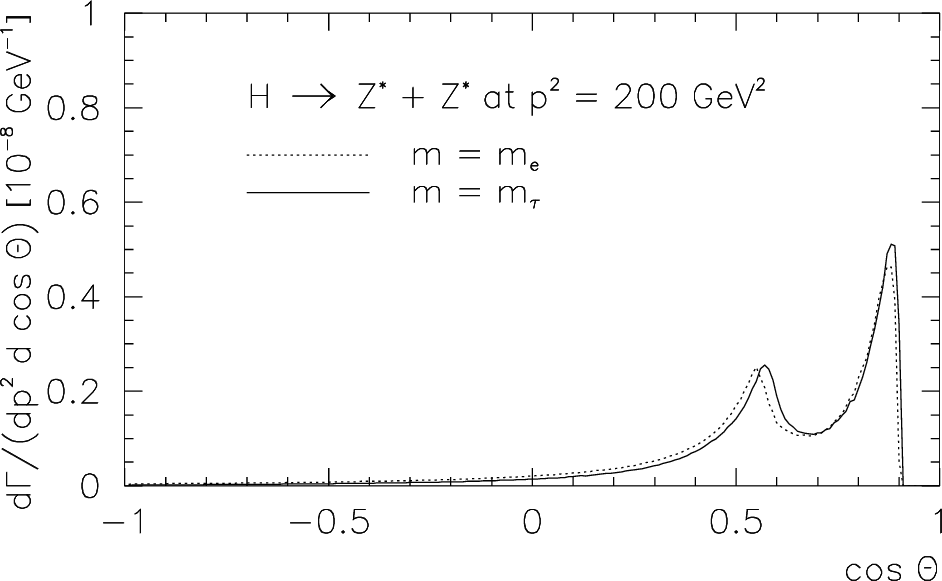,scale=0.8}
\vspace{12pt}
\epsfig{figure=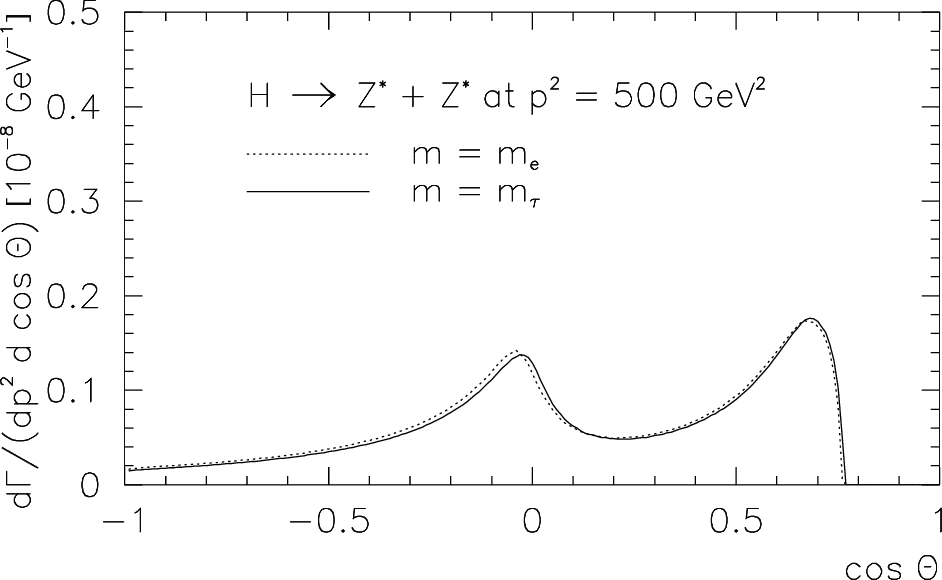,scale=0.8}
\end{center}
\caption{(cont.) Angular dependence of $d\Gamma/(dp^2d\cos\theta)$ for
different squared energies $p^2$}
\end{figure}

\section{Class II contributions}
\begin{figure}\begin{center}
\epsfig{figure=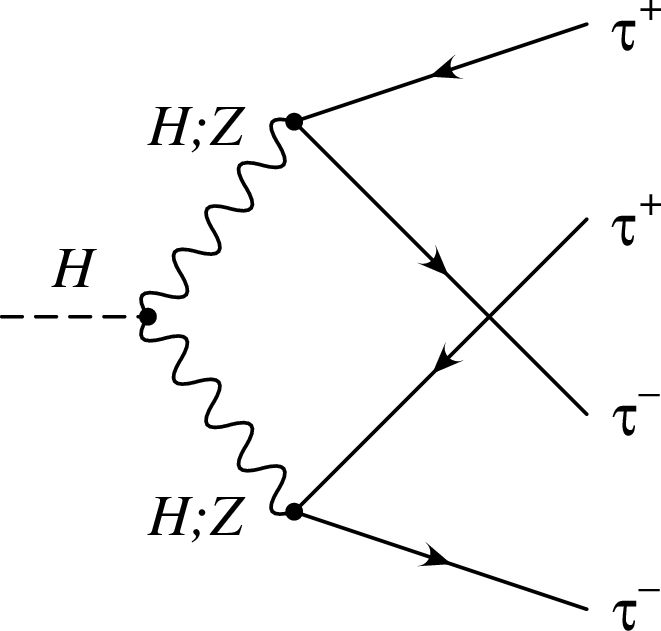, scale=0.3}\qquad
\epsfig{figure=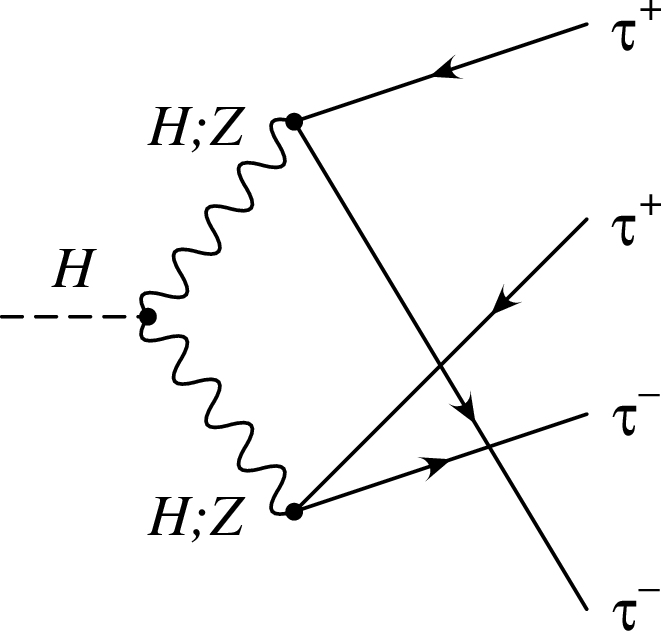, scale=0.3}\qquad
\epsfig{figure=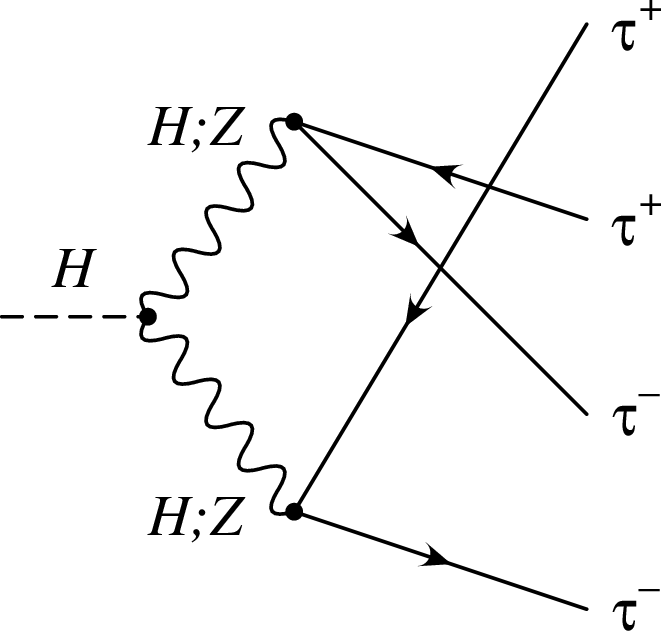, scale=0.3}\qquad
\epsfig{figure=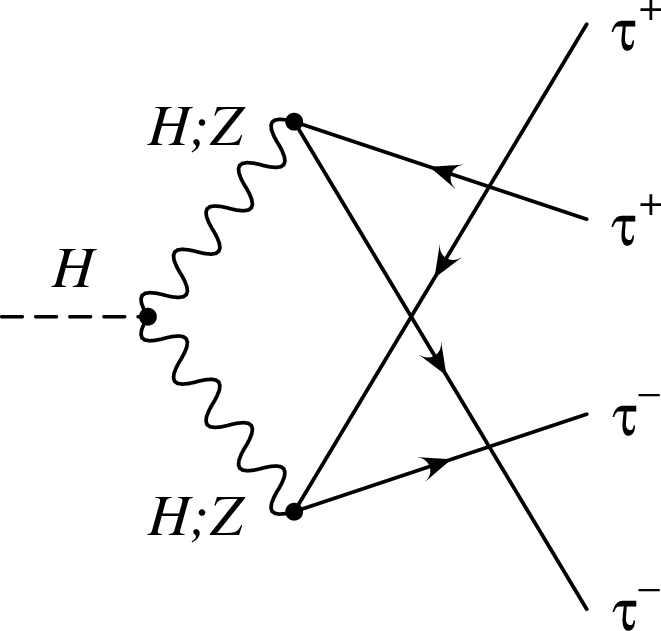, scale=0.3}\\[12pt]
(1a)\kern98pt(1b)\kern98pt(1c)\kern98pt(1d)\\[12pt]
\epsfig{figure=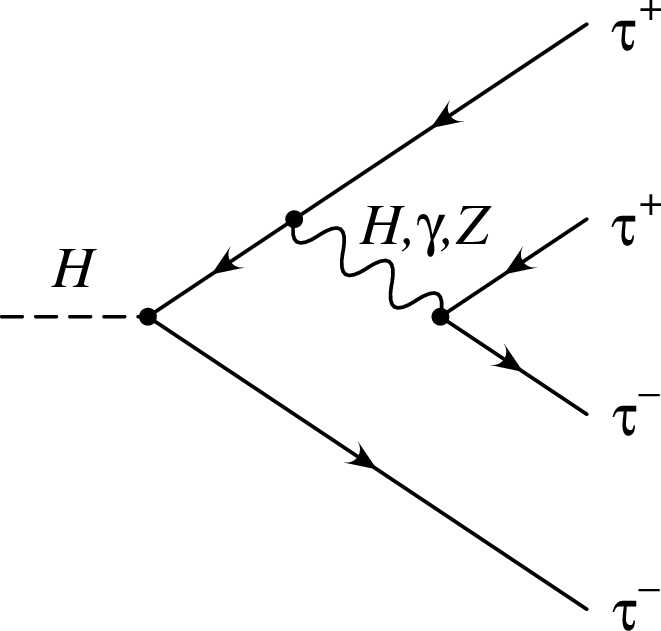, scale=0.3}\qquad
\epsfig{figure=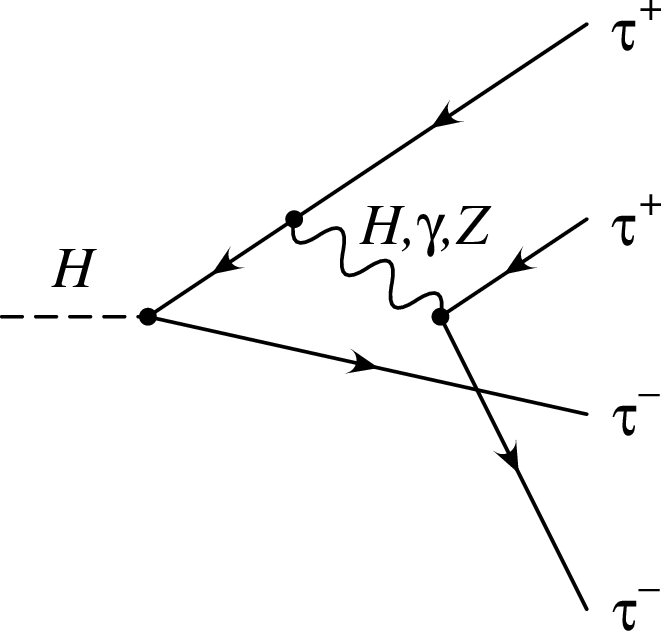, scale=0.3}\qquad
\epsfig{figure=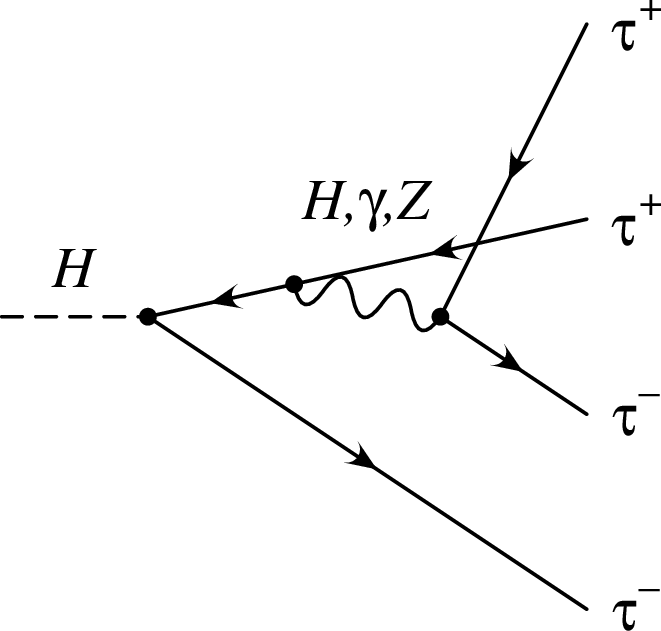, scale=0.3}\qquad
\epsfig{figure=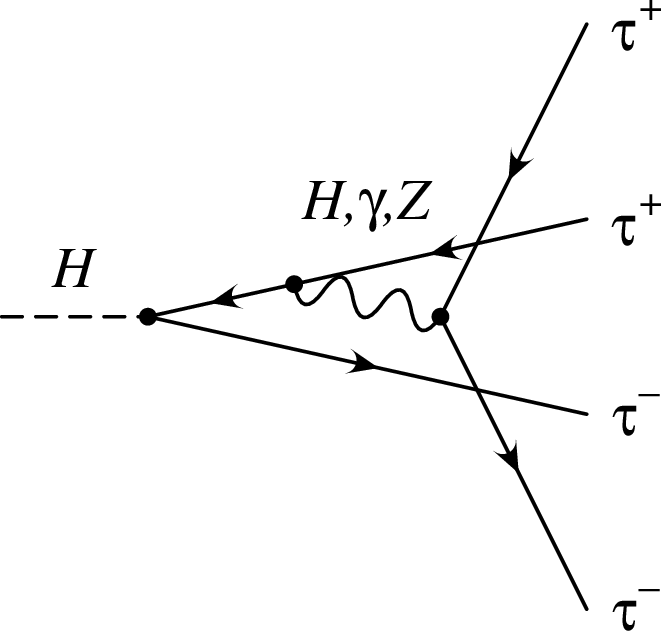, scale=0.3}\\[12pt]
(2a)\kern98pt(2b)\kern98pt(2c)\kern98pt(2d)\\[12pt]
\epsfig{figure=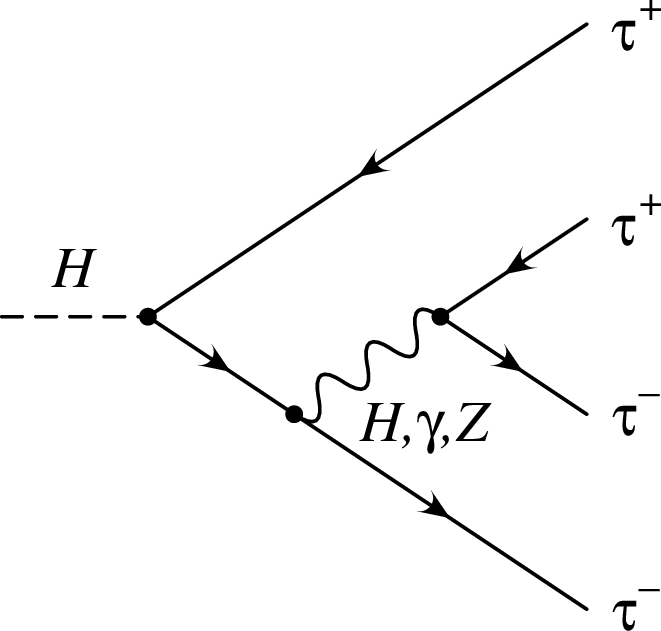, scale=0.3}\qquad
\epsfig{figure=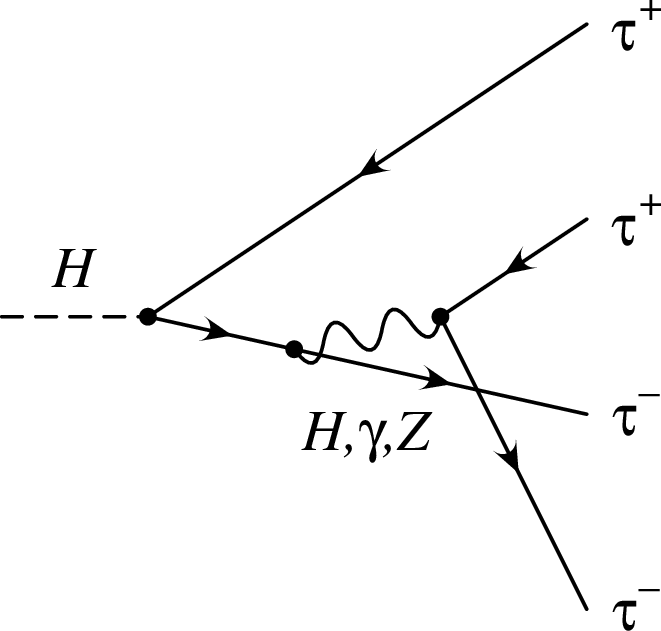, scale=0.3}\qquad
\epsfig{figure=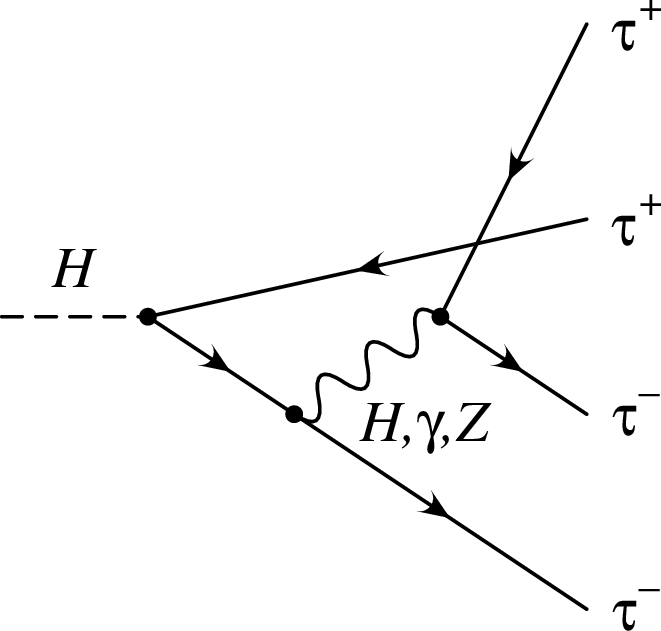, scale=0.3}\qquad
\epsfig{figure=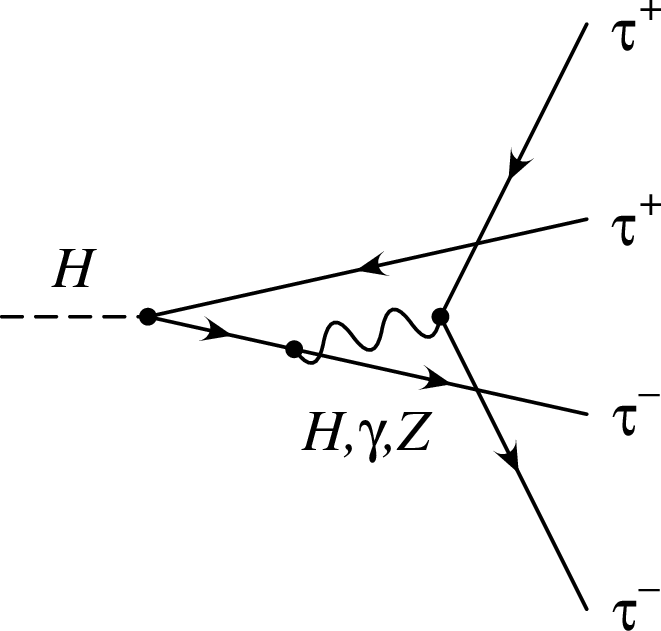, scale=0.3}\\[12pt]
(3a)\kern98pt(3b)\kern98pt(3c)\kern98pt(3d)\\[12pt]
\epsfig{figure=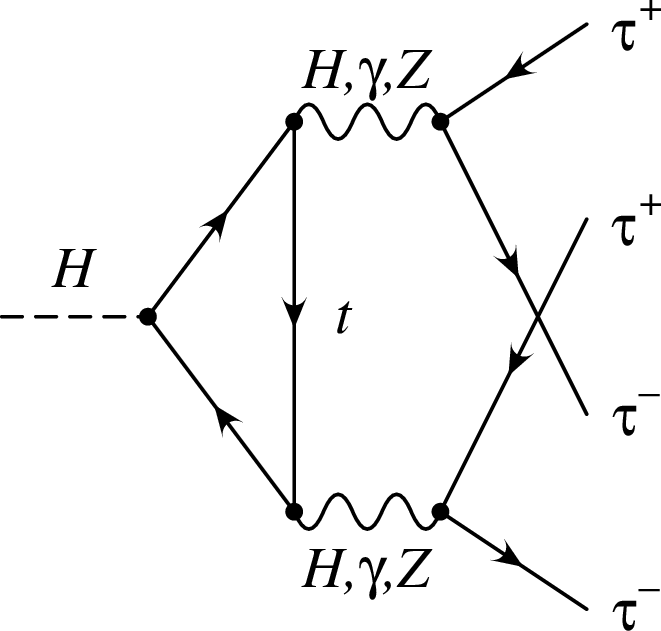, scale=0.3}\qquad
\epsfig{figure=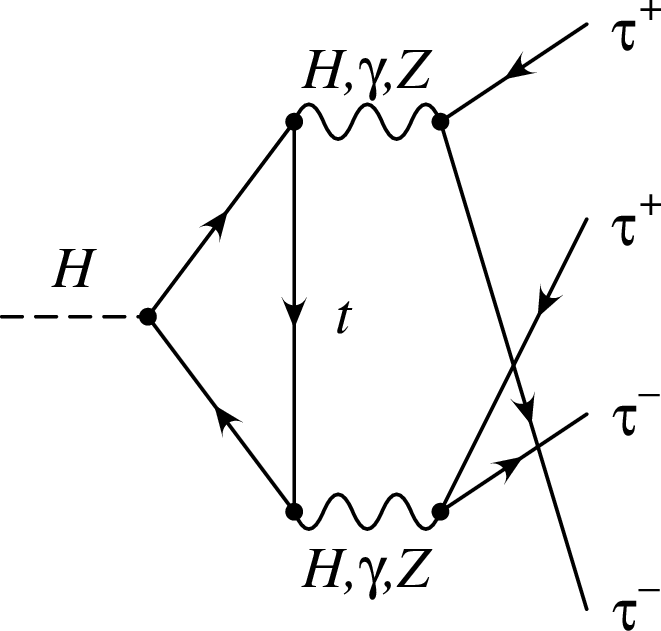, scale=0.3}\qquad
\epsfig{figure=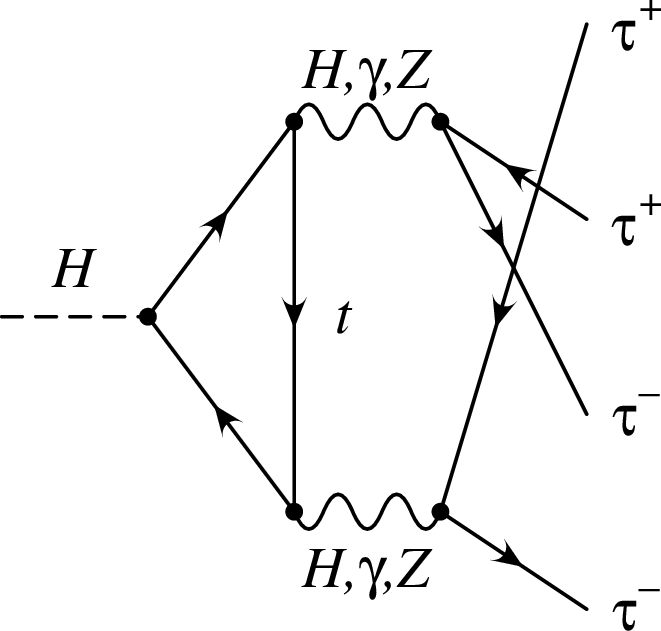, scale=0.3}\qquad
\epsfig{figure=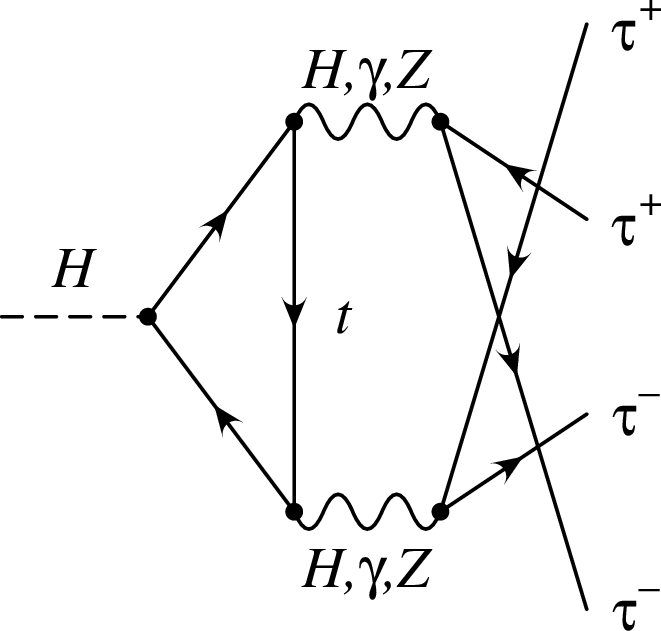, scale=0.3}\\[12pt]
(4a)\kern98pt(4b)\kern98pt(4c)\kern98pt(4d)
\caption{\label{jiandiag}Diagrams contributing to
$H\to\tau^+\tau^+\tau^-\tau^-$}
\end{center}\end{figure}

As emphasized in the Introduction, the class I contributions considered up to
this point and given by the cascade process
$H\to Z^\ast(\to\tau^+\tau^-)+Z^\ast(\to\tau^+\tau^-)$ are not the only
contributions to $H\to\ell\ell\ell\ell$. Instead, because of the finite tau
lepton mass, there are also class II diagrams which were mentioned (though, in
a different context) in Ref.~\cite{Ali:1979rz}. Together with class III
two-loop diagrams we obtain sixteen diagrams (with multiple boson choices)
shown in Fig.~\ref{jiandiag} (outgoing lepton momenta from top to bottom are
$p_1$, $p_2$ $p_3$ and $p_4$). Though, concentrating only on the one-loop
class I and II contributions, the situation is not as hopeless at it might
look at the first sight. The diagrams of the second and third line in
Fig.~\ref{jiandiag} can be combined to give an effective contribution. As the
momentum carried by the intermediate boson line in diagrams~(2a) and~(3a) is
given by $p_2+p_4=p_b$, there are similarities in the kinematics with
diagram~(1a). Denoting the combined diagrams of the second and third line by
$J_i$ and of the first line as before by $M_i$ ($i=a,b,c,d$ -- however,
containing also intermediate Higgs bosons), one obtains four $4\times4$
matrices $J_{ij}=J_i^{\phantom\dagger}J_j^\dagger$,
$K_{ij}=J_i^{\phantom\dagger}M_j^\dagger$,
$L_{ij}=M_i^{\phantom\dagger}J_j^\dagger$ and
$M_{ij}=M_i^{\phantom\dagger}M_j^\dagger$.

\subsection{The contributions $J_{ij}$, $K_{ij}$, $L_{ij}$ and $M_{ij}$}
The different contributions to the absolute square of the matrix element are
given in Tab.~\ref{tabJKLM} where rows and columns stand for the first and
second index, respectively. Calculating the sum over the different squared
decay channels $aa$, $ab$, \ldots divided by the symmetry factor $4$, the
contribution $M+(L+K)+J$ reads
\[(5.89(2)-4.62(7)\cdot 10^{-3}+0.362(2))\cdot 10^{-8}\GeV
  =6.25(2)\cdot 10^{-8}\GeV.\]
If instead of $m=m_\tau=1.77682\GeV$ we choose $m=m_\mu=113.429\MeV$ or
$m=m_e=0.511\MeV$, the contributions of $J_{ij}$, $K_{ij}$ and $L_{ij}$ to
the cross section decrease rapidly. $J_{ij}$ is dominated by the diagonal
elements $i=j$, leading to $1.08\cdot 10^{-10}\GeV$ for $m=m_\mu$ and
$1.4\cdot 10^{-15}\GeV$ for $m_e$. The mixed contributions $K_{ij}$ and
$L_{ij}$ are not dominated by the diagonal elements. These contributions
combined are given by $-1.96\cdot 10^{-13}\GeV$ ($m=m_\mu$) and
$-3.9\cdot 10^{-18}\GeV$ ($m=m_e$). Therefore, in all these cases the
contribution of $M_{ij}$ dominates (cf.\ Tab.~\ref{tabMemt}). For $m=m_\mu$
one obtains
\[(6.12(2)-1.96(5)\cdot 10^{-5}+0.0108(2))\cdot 10^{-8}\GeV
  =6.13(2)\cdot 10^{-8}\GeV,\]
and for $m=m_e$ one obtains
\[(6.12(2)-3.9(1)\cdot 10^{-10}+1.4(1)\cdot 10^{-7})\cdot 10^{-8}\GeV
  =6.12(2)\cdot 10^{-8}\GeV.\]
Therefore, even though the phase space the main contribution $M_{ij}$
increases for decreasing lepton mass, the total cross section decreases due to
the additional diagrams.

\begin{table}\begin{center}
\caption{\label{tabJKLM}Contributions of $J_{ij}$, $K_{ij}$,
$L_{ij}$ and $M_{ij}$ to the total rate for $m=m_\tau$.
The values are given in in units of $10^{-7}\GeV$.}\vspace{12pt}
\begin{tabular}{|c||c|c|c|c|}\hline
$J_{ij}$&$a$&$b$&$c$&$d$\\\hline\hline
$a$&$+0.1440(6)$&$+0.00927(4)$&$+0.00928(4)$&$+0.000837(3)$\\\hline
$b$&$+0.00927(4)$&$+0.1450(2)$&$+0.000836(1)$&$+0.00926(4)$\\\hline
$c$&$+0.00928(4)$&$+0.000836(1)$&$+0.1448(2)$&$+0.00923(4)$\\\hline
$d$&$+0.000837(3)$&$+0.00926(4)$&$+0.00923(4)$&$+0.1447(6)$\\\hline\hline
$K_{ij}$&$a$&$b$&$c$&$d$\\\hline\hline
$a$&$-0.0009(1)$&$-0.0006(1)$&$-0.0006(1)$&$-0.0009(1)$\\\hline
$b$&$-0.0007(1)$&$-0.0009(1)$&$-0.0009(1)$&$-0.0007(1)$\\\hline
$c$&$-0.0007(1)$&$-0.0009(1)$&$-0.0009(1)$&$-0.0007(1)$\\\hline
$d$&$-0.0009(1)$&$-0.0006(1)$&$-0.0006(1)$&$-0.0009(1)$\\\hline\hline
$L_{ij}$&$a$&$b$&$c$&$d$\\\hline\hline
$a$&$-0.0009(1)$&$-0.0007(1)$&$-0.0007(1)$&$-0.0009(1)$\\\hline
$b$&$-0.0006(1)$&$-0.0009(1)$&$-0.0009(1)$&$-0.0006(1)$\\\hline
$c$&$-0.0006(1)$&$-0.0009(1)$&$-0.0009(1)$&$-0.0006(1)$\\\hline
$d$&$-0.0009(1)$&$-0.0007(1)$&$-0.0007(1)$&$-0.0009(1)$\\\hline\hline
$M_{ij}$&$a$&$b$&$c$&$d$\\\hline\hline
$a$&$+2.357(9)$&$+0.2446(8)$&$+0.2446(8)$&$+2.357(9)$\\\hline
$b$&$+0.2446(8)$&$+2.37(2)$&$+2.37(2)$&$+0.2446(8)$\\\hline
$c$&$+0.2446(8)$&$+2.37(2)$&$+2.37(2)$&$+0.2446(8)$\\\hline
$d$&$+2.357(9)$&$+0.2446(8)$&$+0.2446(8)$&$+2.357(9)$\\\hline
\end{tabular}
\end{center}\end{table}

\begin{table}\begin{center}
\caption{\label{tabMemt}Contributions of $M_{ij}$ to the total rate for
different lepton masses $m=m_e,m_\mu,m_\tau$. The values are given in units
  of $10^{-7}\GeV$.}\vspace{12pt}
\begin{tabular}{|c||c|c|c|c|}\hline
$m=m_e$&$a$&$b$&$c$&$d$\\\hline\hline
$a$&$2.447(9)$&$0.2501(8)$&$0.2501(8)$&$2.447(9)$\\\hline
$b$&$0.2501(8)$&$2.46(2)$&$2.46(2)$&$0.2501(8)$\\\hline
$c$&$0.2501(8)$&$2.46(2)$&$2.46(2)$&$0.2501(8)$\\\hline
$d$&$2.447(9)$&$0.2501(8)$&$0.2501(8)$&$2.447(9)$\\\hline\hline
$m=m_\mu$&$a$&$b$&$c$&$d$\\\hline\hline
$a$&$2.446(9)$&$0.2501(8)$&$0.2501(8)$&$2.446(9)$\\\hline
$b$&$0.2501(8)$&$2.46(2)$&$2.46(2)$&$0.2501(8)$\\\hline
$c$&$0.2501(8)$&$2.46(2)$&$2.46(2)$&$0.2501(8)$\\\hline
$d$&$2.446(9)$&$0.2501(8)$&$0.2501(8)$&$2.446(9)$\\\hline\hline
$m=m_\tau$&$a$&$b$&$c$&$d$\\\hline\hline
$a$&$2.357(9)$&$0.2446(8)$&$0.2446(8)$&$2.357(9)$\\\hline
$b$&$0.2446(8)$&$2.37(2)$&$2.37(2)$&$0.2446(8)$\\\hline
$c$&$0.2446(8)$&$2.37(2)$&$2.37(2)$&$0.2446(8)$\\\hline
$d$&$2.357(9)$&$0.2446(8)$&$0.2446(8)$&$2.357(9)$\\\hline
\end{tabular}
\end{center}\end{table}

\section{Conclusions}
In this paper we have studied the decay channel
$H\to Z^\ast(\to\ell^+\ell^-)+Z^\ast(\to\ell^+\ell^-)$ into identical leptons
in detail. We have dealt with the peculiarities of identical particle effects
and worked on the dependence on the lepton mass. We have found that for
increasing lepton mass the decay rates decrease for class-I contributions
while class-II contributions invert this trend. We have shown that nondiagonal
class-I interference contributions are suppressed compared to diagonal class-I
noninterference contributions by about a factor of $10$. Lepton mass dependent
class-II contributions correct the result by $6\%$ but can be safely neglected
for the lighter leptons. Mixed contributions between class-I and class-II
processes can be neglected in all cases. We have dwelled on the narrow width
approximation and we have shown that in the limit of a vanishing vector boson
width $\Gamma_Z$ the nondiagonal interference contribution $\Gamma^{AB}$ stays
constant while the diagonal noninterference contribution $\Gamma^{AA}$ grows,
leading to the approximate narrow width limit
$\Gamma^{AB}/\Gamma^{AA}\to 5(\Gamma_Z/m_Z)$. As a possible observable for
future experiments we worked on single angle decay distributions and
identified the contributions of the diagonal and nondiagonal terms to the two
separate peaks of the distribution, indicating a clear assignment of the
lepton momenta to the intermediate virtual $Z$ bosons or a mixture of those,
respectively.

\subsection*{Acknowledgments}
This work was supported by the Estonian Institutional Research Support under
grant No.~IUT2-27, by the Estonian Science Foundation under grant No.~8769, and
by the European Regional Development Fund under Grant No.~TK133. We would like
to thank A.~Denner, M.~Rauch and J.~Wang for useful discussions and
S.~Dittmaier for hints related to the state-of-the-art Monte Carlo codes
{\sc Prophecy4f} and {\sc Hto4L}. S.G.\ acknowledges the support by the Mainz
Institute of Theoretical Physics (MITP) and by the centers of excellence
PRISMA and PRISMA+. This paper is finished in grateful remembrance on our
deceased colleague J\"urgen G. K\"orner who initiated this research.

\begin{appendix}

\section{Kinematics of the single angle decay distributions}
\setcounter{equation}{0}\def\theequation{A\arabic{equation}}
In this Appendix we deal in detail on the kinematics of the single angle decay
distributions. Of particular interest is the treatment of the delta
distribution in Eq.~(\ref{dGami}) and the construction of the corresponding
integration limits for VEGAS. One has
\begin{eqnarray}\label{pabcd}
p_a^2&=&(p_1+p_3)^2\ =\ p_a^2,\qquad
p_b^2\ =\ (p_2+p_4)^2\ =\ p_b^2,\nonumber\\[3pt]
p_c^2&=&(p_1+p_4)^2\ =\ 2m^2+\frac12\sqrt{p_a^2p_b^2}
  \Big(\cosh(\lambda_a-\lambda_b)+v_a\cos\theta_a\sinh(\lambda_a-\lambda_b)
  +\strut\nonumber\\&&\strut
  +v_b\cos\theta_b\sinh(\lambda_a-\lambda_b)
  +v_av_b\cos\theta_a\cos\theta_b\cosh(\lambda_a-\lambda_b)
  +v_av_b\sin\theta_a\sin\theta_b\cos\phi\Big),\nonumber\\
p_d^2&=&(p_2+p_3)^2\ =\ 2m^2+\frac12\sqrt{p_a^2p_b^2}
  \Big(\cosh(\lambda_a-\lambda_b)-v_a\cos\theta_a\sinh(\lambda_a-\lambda_b)
  +\strut\nonumber\\&&\strut
  -v_b\cos\theta_b\sinh(\lambda_a-\lambda_b)
  +v_av_b\cos\theta_a\cos\theta_b\cosh(\lambda_a-\lambda_b)
  +v_av_b\sin\theta_a\sin\theta_b\cos\phi\Big)\nonumber\\
\end{eqnarray}
where $v_a$, $v_b$, $\lambda_a$ and $\lambda_b$ are given in terms of $m^2$,
$p_a^2$, $p_b^2$ and $m_H^2$. For the function
$\cos\theta^i(p_a^2,p_b^2,\theta_a,\theta_b,\phi)$ we have to take into
account the scalar products and normalizations of the momentum three-vectors,
\begin{equation}
\cos\theta_{ij}=\frac{\vec p_i\cdot\vec p_j}{|\vec p_i||\vec p_j|},
\end{equation}
where for instance $\cos\theta^a=\cos\theta_{13}$ can be constructed with the
help of
\begin{eqnarray}
|\vec p_1|^2&=&-(1+\cos^2\theta_a\sinh^2\lambda_a)m^2+\strut\nonumber\\&&\strut
  +\frac{p_a^2}4\left(\cosh^2\lambda_a+\cos^2\theta_a\sinh^2\lambda_a
  +2v_a\cos\theta_a\sinh\lambda_a\cosh\lambda_a\right),\nonumber\\
|\vec p_3|^2&=&-(1+\cos^2\theta_a\sinh^2\lambda_a)m^2+\strut\nonumber\\&&\strut
  +\frac{p_a^2}4\left(\cosh^2\lambda_a+\cos^2\theta_a\sinh^2\lambda_a
  -2v_a\cos\theta_a\sinh\lambda_a\cosh\lambda_a\right),\nonumber\\
\vec p_1\cdot\vec p_3&=&(1+\cos^2\theta_a\sinh^2\lambda_a)m^2
  -\frac{p_a^2}4(1-\sin^2\theta_a\sinh^2\lambda_a).
\end{eqnarray}
One can use
\begin{equation}
\delta\left(\frac{\vec p_1\cdot\vec p_3}{|\vec p_1||\vec p_3|}-\cos\theta_{13}
  \right)=|\vec p_1||\vec p_3|\delta(\vec p_1\cdot\vec p_3
  -|\vec p_1||\vec p_3|\cos\theta_{13})
\end{equation}
in order to get rid of the ratio in the argument of the delta distribution, and
the square root hidden in the absolute values of the three-vectors can be
removed by using
\begin{eqnarray}
\lefteqn{\delta\left((\vec p_1\cdot\vec p_3)^2
  -|\vec p_1|^2|\vec p_3|^2\cos^2\theta_{13}\right)
  \ =\ \frac1{2|\vec p_1||\vec p_3||\vec p_1\cdot\vec p_3|}
  \strut}\nonumber\\&&\strut\times
  \left(\delta\left(\frac{\vec p_1\cdot\vec p_3}{|\vec p_1||\vec p_3|}
  -\cos\theta_{13}\right)
  +\delta\left(\frac{\vec p_1\cdot\vec p_3}{|\vec p_1||\vec p_3|}
  +\cos\theta_{13}\right)\right).
\end{eqnarray}
The dublication of zeros corresponds to a multiplicity of solutions for the
zeroth of the argument of the delta distribution. Solving
$(\vec p_1\cdot\vec p_3)^2-|\vec p_1|^2|\vec p_3|^2\cos^2\theta_{13}=0$
for $\cos\theta_a$ leads to four solutions
\begin{equation}
c_a=\pm\sqrt{\frac{(p_a^2\cosh^2\lambda_a+4m^2)\sin^2\theta_{13}
  -2p_a^2\pm\sqrt{4p_a^2\cos^2\theta_{13}(p_a^2-4m^2\cosh^2\lambda_a
  \sin^2\theta_{13})}}{(p_a^2-4m^2)\sinh^2\lambda_a\sin^2\theta_{13}}}
\end{equation}
which can be written symbolically as
\begin{equation}\label{branches}
c_a^{(a)}=\sqrt{\frac{N_a+R_a}{D_a}},\quad
c_a^{(b)}=\sqrt{\frac{N_a-R_a}{D_a}},\quad
c_a^{(c)}=-\sqrt{\frac{N_a+R_a}{D_a}}\quad\mbox{and}\quad
c_a^{(d)}=-\sqrt{\frac{N_a-R_a}{D_a}}.
\end{equation}
\begin{figure}
\epsfig{figure=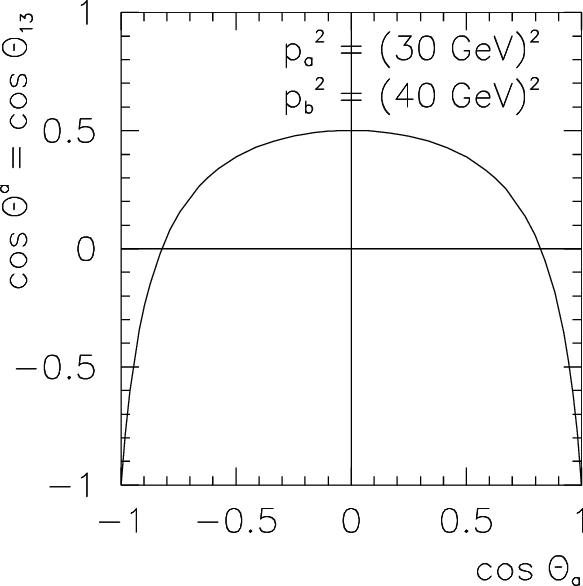, scale=0.7}\qquad
\epsfig{figure=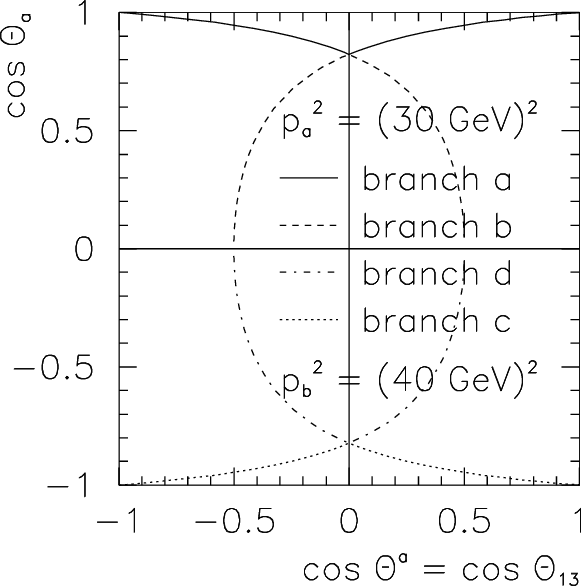, scale=0.7}
\caption{\label{ca13abcd}function $\cos\theta_{13}(\cos\theta_a)$ (left panel)
and the four branches $a$, $b$, $c$ and $d$ in Eq.~(\ref{branches}) for the
inverted function $\cos\theta_a(\cos\theta_{13})$ (right panel) at
$\sqrt{p_a^2}=30\GeV$ and $\sqrt{p_b^2}=40\GeV$}
\end{figure}
In Fig.~\ref{ca13abcd} we compare the original function
$\cos\theta_{13}(\cos\theta_a)$ (left panel) with the four branches $a$, $b$,
$c$ and $d$ for the inverted function $\cos\theta_a(\cos\theta_{13})$ (right
panel). Obviously, only half of the branches have to be taken, leading to two
solutions (zeros)
\begin{equation}
c_a^+=\cases{c_a^{(a)}&for $\cos\theta_{13}\le 0$,\cr
  c_a^{(b)}&for $\cos\theta_{13}\ge 0$,\cr},\qquad
c_a^-=\cases{c_a^{(c)}&for $\cos\theta_{13}\le 0$,\cr
  c_a^{(d)}&for $\cos\theta_{13}\ge 0$,\cr}
\end{equation}
which can be witten again in the compact form
\begin{equation}
c_a^\pm=\pm\sqrt{\frac{(p_a^2\cosh^2\lambda_a+4m^2)\sin^2\theta_{13}
  -2p_a^2-2\cos\theta_{13}\sqrt{p_a^2(p_a^2-4m^2\cosh^2\lambda_a
  \sin^2\theta_{13})}}{(p_a^2-4m^2)\sinh^2\lambda_a\sin^2\theta_{13}}}
\end{equation}
Employing the known rule $\delta(f(x))=\sum_i\delta(x-x_i)/|f'(x_i)|$ where
the sum runs over the zeros $x_i$ of $f(x)$, each term in the sum of delta
distributions will lead to the replacement of $\cos\theta_a$ by the
corresponding zero $c_a^\pm$. After a long calculation involving a lot of
simplifications by hand (but checked afterwards numerically), we obtain
\begin{equation}
  \delta\left(\frac{\vec p_1\cdot\vec p_3}{|\vec p_1||\vec p_3|}
  -\cos\theta_{13}\right)
  =\frac1{C_a}\left(\delta(\cos\theta_a-c_a^+)+\delta(\cos\theta_a-c_a^-)
  \right),
\end{equation}
where
\begin{eqnarray}\label{cera}
C_a&:=&v_a\sinh\lambda_a\frac{\sqrt{r_a}n_a}{2t_a^{3/2}},\nonumber\\
r_a&:=&\left(p_a^2\sinh^2\lambda_a+4m^2(1-\cosh^2\lambda_a\cos^2\theta_{13})
  \right)\sin^2\theta_{13}-q_a,\nonumber\\[3pt]
n_a&:=&\left(p_a^2-4m^2\cosh^2\lambda_a(2-\cos^2\theta_{13})\right)
  \sin^2\theta_{13}+q_a,\nonumber\\
t_a&:=&\frac{q_a}{\sin^2\theta_{13}}-4m^2\cosh^2\lambda_a\sin^2\theta_{13},
  \nonumber\\
q_a&:=&\left(\sqrt{p_a^2}+\cos\theta_{13}\sqrt{p_a^2-4m^2\cosh^2\lambda_a
  \sin^2\theta_{13}}\right)^2,\nonumber\\[3pt]
d_a&:=&(p_a^2-4m^2)\sinh^2\lambda_a\sin^2\theta_{13}.
\end{eqnarray}
It is easy to see that $c_a^\pm=\pm\sqrt{r_a/d_a}$. A corresponding explicit
relation is possible between $\theta_{24}$ and $c_b$, but an analytic
resolution fails for $\theta_{14}$ and $\theta_{23}$.

\subsection{Constraints for the angle $\theta_{13}$}
\begin{figure}
\begin{center}
\epsfig{figure=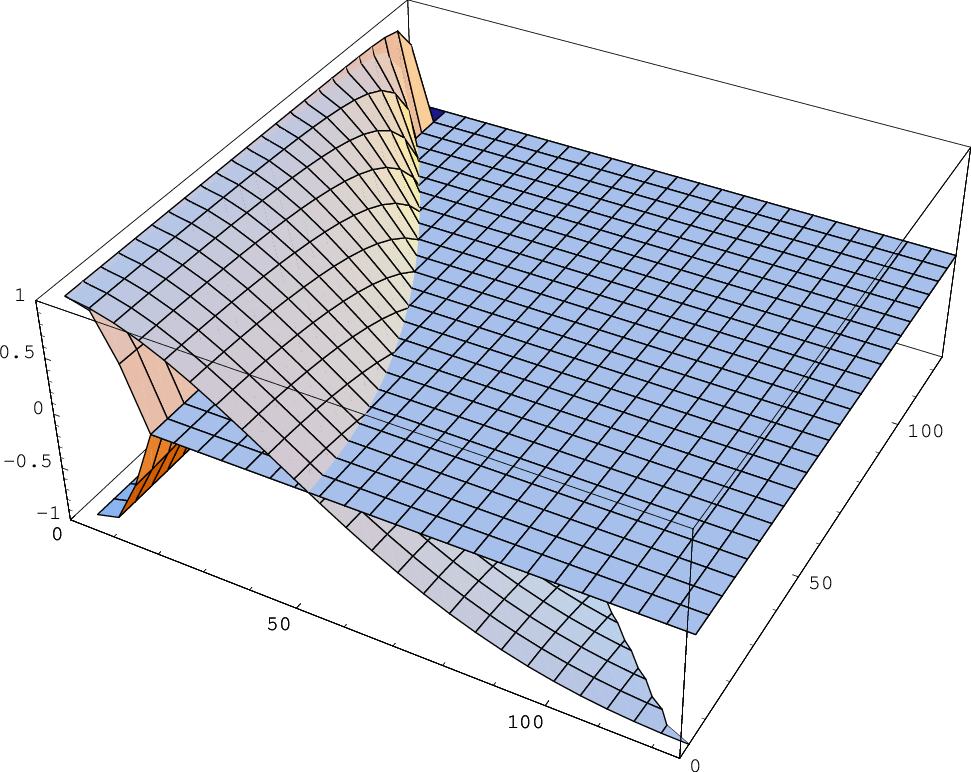, scale=0.8}
\caption{\label{c13mia}Constraints on $\cos\theta_{13}$ (to the top) in
dependence on $\sqrt{p_a^2}$ (in $\GeV$, to the right) and $\sqrt{p_b^2}$
(in $\GeV$, to the back). The plane continuation at $\cos\theta_{13}=0$ has to
be ignored.}
\end{center}
\end{figure}

\begin{figure}
\begin{center}
\epsfig{figure=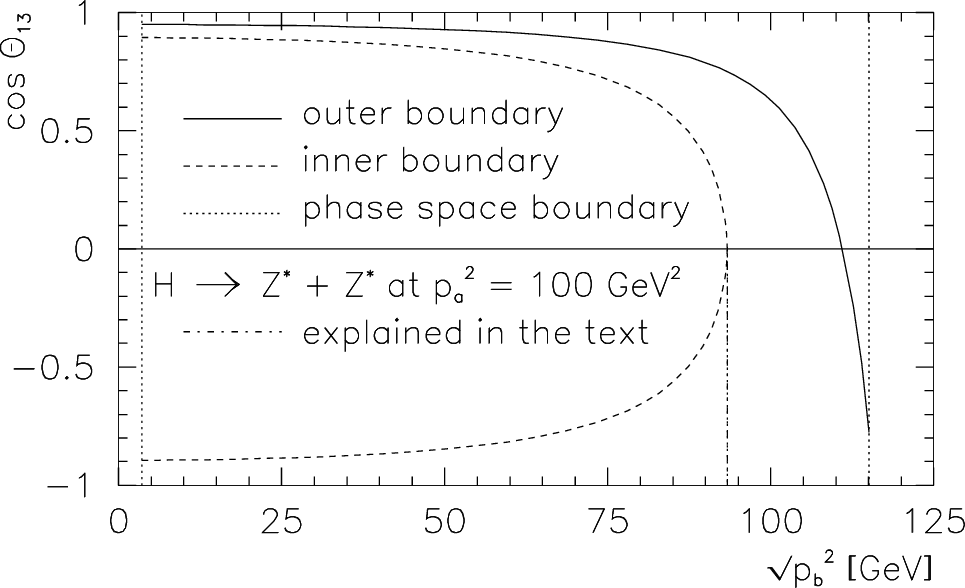, scale=0.8}
\caption{\label{c13mia100}Constraints on $\cos\theta_{13}$ in dependence on
$\sqrt{p_b^2}$ for $p_a^2=100\GeV^2$}
\end{center}
\end{figure}

In performing the calculation with VEGAS, the calculation experiences a
couple of branching points, i.e.\ phase space values have to be excluded where
either the radicant
\begin{equation}
p_a^2-4m^2\cosh^2\lambda_a\sin^2\theta_{13}
\end{equation}
in $q_a$ or the radicant $r_a$ of the final result for $c_a^\pm$ becomes
negative. The restriction to the former one imposes the minimum condition
\begin{equation}
|\cos\theta_{13}|\ge\sqrt{1-\frac{p_a^2}{4m^2\cosh^2\lambda_a}},
\end{equation}
while the restriction to the latter one leads to the maximum condition
\begin{equation}
\cos\theta_{13}\le\frac{p_a^2(\sinh^2\lambda_a-1)+4m^2}{p_a^2
  (\sinh^2\lambda_a+1)+4m^2}.
\end{equation}
Both constraints are displayed in Fig.~\ref{c13mia}, leaving a narrow region
where $\cos\theta_{13}$ is defined. In Fig.~\ref{c13mia100} we cut these
surfaces of constraint at $p_a^2=100\GeV^2$. The region is narrow in
particular for small $p_a^2$, explaining the pile-up of the peak close to the
upper boundary shown in Fig.~\ref{mdpsxxx}.

\begin{figure}
\epsfig{figure=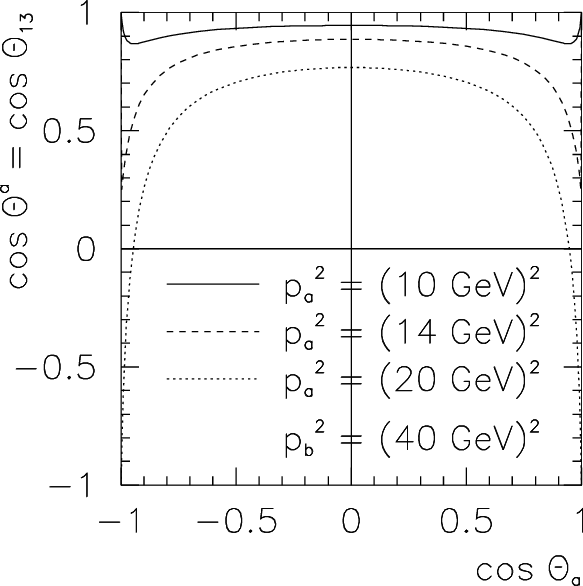, scale=0.7}\qquad
\epsfig{figure=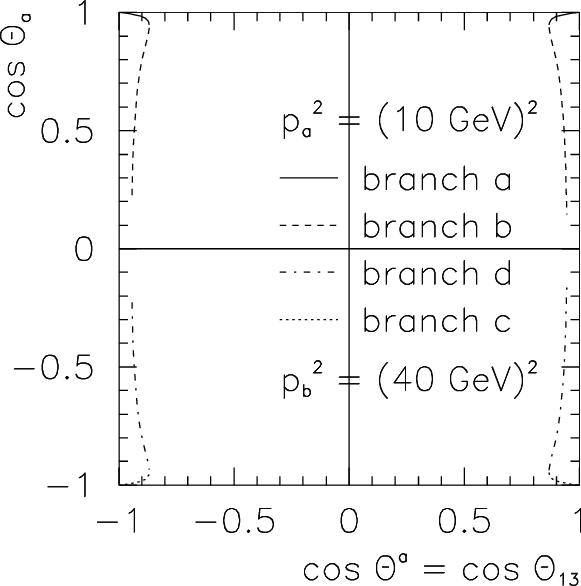, scale=0.7}
\caption{\label{cb13abcd}function $\cos\theta_{13}(\cos\theta_a)$ (left hand
side) and the four branches $a$, $b$, $c$ and $d$ in Eq.~(\ref{branches})
for the inverted function $\cos\theta_a(\cos\theta_{13})$ (right hand side)
at $\sqrt{p_a^2}=10\GeV$ and $\sqrt{p_b^2}=40\GeV$}
\end{figure}

Note that the dependence of the opening angle $\theta^a=\theta_{13}$ between
the two leptons on the angle $\theta_a$ of the leptons with respect to the
momentum direction of the $Z$ boson shown in the left panel of
Fig.~\ref{ca13abcd} can be understood physically: If $\theta_a=0$ or
$\theta_a=\pi$, i.e.\ the two $\tau$ leptons are emitted in the boost
direction, the boosted momenta remain in this direction. The maximal effect is
found for $\theta_a=\pi/2$ where the two $\tau$ leptons are emitted
perpendicularly to the $Z$ momentum direction. In this case the two momenta
are boosted to the front. However, one has to be careful about joining the
branches. If channel $a$ is quite ``light'' (i.e.\ $p_a^2$ small), the boost
is close to the light cone and the $\tau$ lepton emitted in back direction
will be boosted to the front as well. This is shown in the left panel of
Fig.~\ref{cb13abcd} for decreasing values $\sqrt{p_a^2}=20\GeV$, $14\GeV$ and
$10\GeV$. In case of $\sqrt{p_a^2}=10\GeV$ as in the right panel of
Fig.~\ref{cb13abcd} one cannot combine two branches to obtain a function but
remains with four distrinct branches which have to be considered for
$\cos\theta_{13}>0$. The situation flips for $\sinh\lambda_a<v_a\cosh\lambda_a$
or
\begin{equation}
p_a^2<m\frac{m_h^2-p_b^2}{m_h-m}\quad\Leftrightarrow\quad
p_b^2<m_h^2-\frac{m_h-m}{m}p_a^2
\end{equation}
which is a pure mass effect.
The flip point can be found in Fig.~\ref{c13mia100} as the apex of the
inverted parabola. This means that values with negative values of
$\cos\theta_{13}$ below this threshold have to be omitted.\footnote{Otherwise,
for small values of $\sqrt{p_a^2}$ one would obtain a small cusp close to
$\cos\theta_{13}=-1$.} Instead, one has to take into account all four
branches, as being obvious from the right panel of Fig.~\ref{cb13abcd},
leading to the modified result
\begin{eqnarray}
\lefteqn{\delta\left(\frac{\vec p_1\cdot\vec p_3}{|\vec p_1||\vec p_3|}
  -\cos\theta_{13}\right)=\Theta(\cos\theta_{13})\times}\\&&\times
  \left[\frac1{2C_a}\left(\delta(\cos\theta_a-c_a^{(a)})
  +\delta(\cos\theta_a-c_a^{(c)})\right)+\frac1{2\bar C_a}
  \left(\delta(\cos\theta_a-c_a^{(b)})+\delta(\cos\theta_a-c_a^{(d)})\right)
  \right],\nonumber
\end{eqnarray}
where (for $\cos\theta_{13}>0$ which is expressed by the Heaviside step
function) one has $c_a^{(a)}=c_a^+$, $c_a^{(c)}=c_a^-$ and
$c_a^{(b)}=\bar c_a^+$, $c_a^{(d)}=\bar c_a^-$ with
$\bar c_a^\pm=\pm\sqrt{\bar r_a/d_a}$, $\bar C_a$ and $\bar r_a$ calculated
by replacing $q_a$ by
\begin{equation}
\bar q_a\ :=\ \left(\sqrt{p_a^2}-\cos\theta_{13}\sqrt{p_a^2
  -4m^2\cosh^2\lambda_a\sin^2\theta_{13}}\right)^2,
\end{equation}
i.e.\ by formally replacing $\cos\theta_{13}\to-\cos\theta_{13}$. However,
this second integration region can again be omitted, because for
$p_a^2<4m^2\cosh^2\lambda_a$ and $\cos\theta_{13}<0$ the expression $\bar t_a$
is negative, as
\begin{eqnarray}
p_a^2-4m^2\cosh^2\lambda_a(1-\cos^2\theta_{13})
  &<&4m^2\cosh^2\lambda_a\cos^2\theta_{13},\nonumber\\[7pt]
\sqrt{p_a^2}-\cos\theta_{13}\sqrt{p_a^2-4m^2\cosh^2\lambda_a
  (1-\cos^2\theta_{13})}&<&2m\cosh\lambda_a\sin^2\theta_{13},\nonumber\\[7pt]
\bar q_a=\left(\sqrt{p_a^2}-\cos\theta_{13}\sqrt{p_a^2-4m^2\cosh^2\lambda_a
  (1-\cos^2\theta_{13})}\right)^2&<&4m^2\cosh^2\theta_{13}\sin^4\theta_{13}
  \qquad
\end{eqnarray}
and, therefore,
\begin{equation}
\bar t_a=\frac{q_a}{\sin^2\theta_{13}}-4m^2\cosh^2\lambda_a\sin^2\theta_{13}<0.
\end{equation}
Drawing a straight line from the apex in Fig~\ref{c13mia100} to the bottom of
the diagram and skipping the region left to this line, one ends up with the
integration region that is purely functional.

\subsection{Constraints for the angle $\theta_{14}$}
\begin{figure}
\begin{center}
\epsfig{figure=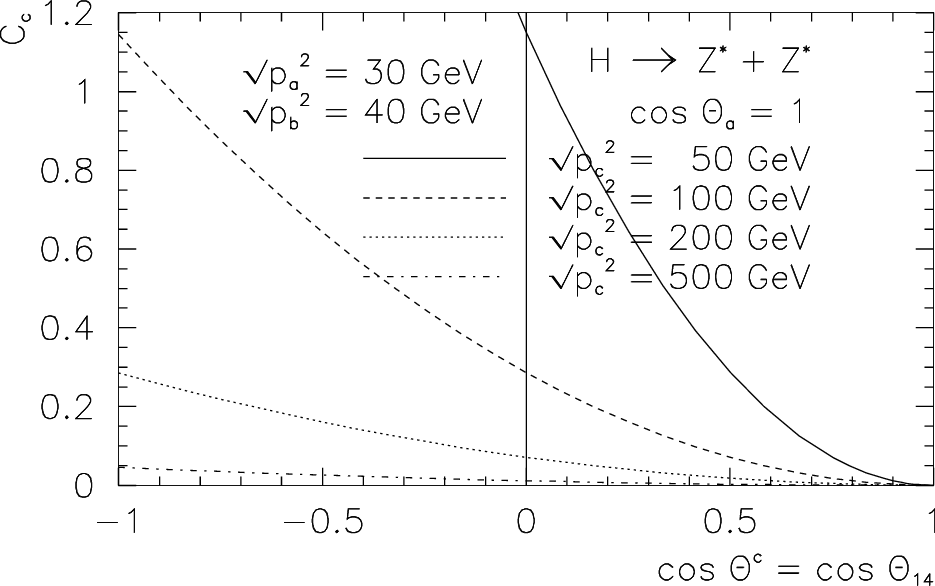,scale=0.8}
\caption{\label{dfun14cp}function $C_c(\cos\theta^c)$ for $\cos\theta_a=1$
and different values for the energy $\sqrt{p_c^2}$}
\end{center}
\end{figure}

Though the method explained before does not work for $\theta^c=\theta_{14}$
and $\theta^d=\theta_{23}$, there is a possibility to solve the delta
distributions also in these cases. For this note that the momentum square
equations for $p_c^2$ and $p_d^2$ in Eqs.~(\ref{pabcd}) can be solved for
$\cos\phi$ almost trivially, inserted into the angular cosine equations, and
solved analytically for $\cos\theta_a$ (or $\cos\theta_b$). As the procedure
is quite similar for the $c$ and $d$ channels, the results are written down
for the $c$ channel only. Solving for $\cos\phi$, one obtains the solution
$\cos\phi_0$ with
\begin{eqnarray}
\lefteqn{v_av_b\sin\theta_a\sin\theta_b\cos\phi_0\ =\ 
  \frac{2(p_c^2-2m^2)}{\sqrt{p_a^2p_b^2}}+\strut}\nonumber\\&&
  -(1+v_av_b\cos\theta_a\cos\theta_b)\cosh(\lambda_a-\lambda_b)
  -(v_a\cos\theta_a+v_b\cos\theta_b)\sin(\lambda_a-\lambda_b)
\end{eqnarray}
and, therefore,
\begin{equation}
\delta\left(p_c^2(p_a^2,p_b^2,\theta_a,\theta_b,\phi)-p_c^2\right)
  =\frac{2(p_a^2p_b^2)^{-1/2}}{v_av_b|\sin\theta_a\sin\theta_b\sin\phi_\phi|}
  \delta(\phi_0-\phi).
\end{equation}
There seems to be a risk that $\theta_a$, $\theta_b$, $v_a$ or $v_b$ might
vanish. Kinematically, this is the case if one of the decay planes is no
longer spanned up and, therefore, the relative angle $\phi$ is not given. If
the expressions to which we insert $\cos\phi$ also contain these factors, one
can assign an arbitrary value to $\phi$, but if this is not the case, these
points have to be skipped artificially for the numerical integration by VEGAS.
A hint for the first case is that if we insert $\cos\phi$ into the second
equation, these factors cancel out. Solving for $\cos\theta_b$ one obtains
\begin{equation}
\cos\theta_b=\frac{N^c_b\pm c_\theta R^c_b}{D^c_b}=:c_b^\pm
\end{equation}
(note the absence of a general square root) with
\begin{eqnarray}
D^c_b&=&p_a^2p_b^2v_b^2\sinh^2\lambda_b\Big[c_a^2-c_\theta^2\Big],\\[7pt]
N^c_b&=&\sqrt{p_a^2p_b^2}v_b\sinh\lambda_bc_a(2p_c^2-4m^2)
  -p_a^2p_b^2v_b\sinh\lambda_b\cosh\lambda_b\left(c_a^2-c_\theta^2\right),
  \\[7pt]
R^c_b&=&\sqrt{p_a^2p_b^2}v_b\sinh\lambda_b
  \sqrt{(2p_c^2-4m^2)^2-4m^2p_a^2(c_a^2-c_\theta^2)},\qquad
\end{eqnarray}
where we have used the abbreviation
$c_\theta:=\cos\theta_{14}\sqrt{s_a^2+v_a^2\sin^2\theta_a}$ as well as
$c_a:=\cosh\lambda_a+v_a\cos\theta_a\sinh\lambda_a$ and
$s_a:=\sinh\lambda_a+v_a\cos\theta_a\cosh\lambda_a$ in order to obtain a
compact form. Also in this case the calculation of $|f'(x_i)|$ and,
therefore, $C_c$ was successful. Analysing the pole structure as before, it
can be seen that only $c_b^+$ can be considered as a relevant zero. One
obtains
\begin{equation}
\delta\left(\frac{\vec p_1\cdot\vec p_4}{|\vec p_1||\vec p_4|}
  -\cos\theta_{14}\right)=\frac1{C_c}\delta(c_b^+-\cos\theta_b),
\end{equation}
where
\begin{equation}
C_c=\frac{\sqrt{p_a^2p_b^2}v_b\sinh\lambda_b(c_a^2-c_\theta^2)^2
  n_c}{\sqrt{s_a^2+v_a^2\sin^2\theta_a}d_c^3}
\end{equation}
with
\begin{eqnarray}
n_c&:=&c_a\left((2p_c^2-4m^2)^2-4m^2p_a^2(c_a^2-c_\theta^2)\right)
  +\strut\nonumber\\&&\strut
  +c_\theta(2p_c^2-4m^2)\sqrt{(2p_c^2-4m^2)^2
  -4m^2p_a^2(c_a^2-c_\theta^2)},\nonumber\\[7pt]
d_c&:=&c_\theta(2p_c^2-4m^2)+c_a\sqrt{(2p_c^2-4m^2)^2
  -4m^2p_a^2(c_a^2-c_\theta^2)}.
\end{eqnarray}
The function $C_c(\cos\theta^c)$ shown in Fig.~\ref{dfun14cp} for different
energies $\sqrt{p_c^2}$ is quite similar to the function $C_a(\cos\theta^a)$.
Therefore, we expect similar angular distributions.

The results obtained so far lead to the constraints we are looking for. The
first condition is
\begin{equation}
(2p_c^2-4m^2)^2-4m^2p_a^2(c_a^2-c_\theta^2)\ge 0\quad\Leftrightarrow\quad
\cos^2\theta_{14}\ge\frac1{s_a^2+v_a^2\sin^2\theta_a}
  \left(c_a^2-\frac{(2p_c^2-4m^2)^2}{4m^2p_a^2}\right).
\end{equation}
On the other hand, the condition $c_b^+\le 1$ leads to
$\cos^2\theta_{14}\le c_a^2/(s_a^2+v_a^2\cos^2\theta_a)$ which is satisfied
trivially, and
\begin{equation}
\cos^2\theta_{14}\le\frac{((2p_c^2-4m^2)-\sqrt{p_a^2p_b^2}c_ac_b)^2}{p_a^2
  p_b^2(s_a^2+v_a^2\sin^2\theta_a)s_b^2}
\end{equation}
with $s_b:=\sinh\lambda_b+v_b\cosh\lambda_b$ and
$c_b:=\cosh\lambda_b+v_b\sinh\lambda_b$. This is a quite weak condition,
working only close to $\cos\theta_a=-1$ (anticollinear lepton with momentum
$p_1$).

\section{Comment on the class III contributions}
\setcounter{equation}{0}\def\theequation{B\arabic{equation}}
In this publication, we do not afford a complete analysis of class III
contributions. However, in order to estimate the order of magnitude of the
contributions, we have calculated two exemplary contributions. The first
contribution we present here is the calculation of the exemplary diagram shown
in Fig.~\ref{htauIII}, i.e., the diagram including the $H\gamma\gamma$ vertex
via a top quark loop. The vertex factor can be expressed in terms of the UV
finite parts of the two- and three-point functions and reads
($\lambda=\lambda(m_H^2,p_a^2,p_b^2)$)
\begin{eqnarray}
\lefteqn{V_{HAA}^{\mu\nu}\ =\ \frac{iem_Z}{c_Ws_W}\times
\frac{m_t^2}{m_Z^2}\times\frac{\alpha Q_t^2}{4\pi\lambda^2}
  \Big\{\hat C_0^f\Big[\lambda\left(4m_H^2p_a^2p_b^2-(m_H^2-p_a^2-p_b^2
  -4m_t^2)\lambda\right)g^{\mu\nu}+\strut}\nonumber\\&&\strut\qquad
  +4p_b^2\left(12m_H^2p_a^2p_b^2+(m_H^2+p_a^2+p_b^2+4m_t^2)\lambda\right)
  p_a^\mu p_a^\nu+\strut\nonumber\\&&\strut\qquad
  -2(m_H^2-p_a^2-p_b^2)\left(12m_H^2p_a^2p_b^2+(m_H^2+p_a^2+p_b^2+4m_t^2)
  \lambda\right)p_a^\mu p_b^\nu+\strut\nonumber\\&&\strut\qquad
  -2(m_H^2-p_a^2-p_b^2)\left(12m_H^2p_a^2p_b^2-(m_H^2-p_a^2-p_b^2-4m_t^2)
  \lambda\right)p_b^\mu p_a^\nu+\strut\nonumber\\&&\strut\qquad
  +4p_a^2\left(12m_H^2p_a^2p_b^2+(m_H^2+p_a^2+p_b^2+4m_t^2)\lambda\right)
  p_b^\mu p_b^\nu\Big]+\strut\nonumber\\&&\strut
  +2\hat B_0^f\Big[\lambda\left(m_H^2(m_H^2-p_a^2-p_b^2)-\lambda\right)
  g^{\mu\nu}
  +4p_b^2\left(3m_H^2(m_H^2-p_a^2-p_b^2)-\lambda\right)p_a^\mu p_a^\nu
  +\strut\nonumber\\&&\strut\qquad
  -2(m_H^2-p_a^2-p_b^2)\left(3m_H^2(m_H^2-p_a^2-p_b^2)-\lambda\right)
  p_a^\mu p_b^\nu+\strut\nonumber\\&&\strut\qquad
  -2\left(12m_H^2p_a^2p_b^2+(p_a^2+p_b^2)\lambda\right)p_b^\mu p_a^\nu
  +4p_a^2\left(3m_H^2(m_H^2-p_a^2-p_b^2)-\lambda\right)p_b^\mu p_b^\nu\Big]
  +\strut\nonumber\\&&\strut
  -2\hat B_a^f\Big[p_a^2\lambda(m_H^2-p_a^2+p_b^2)g^{\mu\nu}
  +4p_b^2\left(3p_a^2(m_H^2-p_a^2+p_b^2)+\lambda\right)p_a^\mu p_a^\nu
  +\strut\nonumber\\&&\strut\qquad
  -2(m_H^2-p_a^2-p_b^2)\left(3p_a^2(m_H^2-p_a^2+p_b^2)+\lambda\right)
  p_a^\mu p_b^\nu+\strut\nonumber\\&&\strut\qquad
  -2p_a^2\left(6p_b^2(m_H^2+p_a^2-p_b^2)+\lambda\right)p_b^\mu p_a^\nu
  +4p_a^2\left(3p_a^2(m_H^2-p_a^2+p_b^2)+\lambda\right)p_b^\mu p_b^\nu\Big]
  +\strut\nonumber\\&&\strut
  -2\hat B_b^f\Big[p_b^2\lambda(m_H^2+p_a^2-p_b^2)g^{\mu\nu}
  +4p_b^2\left(3p_b^2(m_H^2+p_a^2-p_b^2)+\lambda\right)p_a^\mu p_a^\nu
  +\strut\nonumber\\&&\strut\qquad
  -2(m_H^2-p_a^2-p_b^2)\left(3p_b^2(m_H^2+p_a^2-p_b^2)+\lambda\right)
  p_a^\mu p_b^\nu+\strut\nonumber\\&&\strut\qquad
  -2p_b^2\left(6p_a^2(m_H^2-p_a^2+p_b^2)+\lambda\right)p_b^\mu p_a^\nu
  +4p_a^2\left(3p_b^2(m_H^2+p_a^2-p_b^2)+\lambda\right)p_b^\mu p_b^\nu\Big]
  \Big\},\qquad
\end{eqnarray}
where
\begin{eqnarray}
B_0&=&\frac{i\bar\mu^{-2\eps}}{(4\pi)^2}\left[\frac1\eps
  -\ln\pfrac{m_t^2}{\bar\mu^2}+\hat B_0^f\right],\quad
  \hat B_0^f=2\sqrt{\frac{4m_t^2}{q^2}-1}
  \arctan\left(\sqrt{\frac{4m_t^2}{q^2}-1}\right),\nonumber\\
B_a&=&\frac{i\bar\mu^{-2\eps}}{(4\pi)^2}\left[\frac1\eps
  -\ln\pfrac{m_t^2}{\bar\mu^2}+\hat B_a^f\right],\quad
  \hat B_a^f=2\sqrt{\frac{4m_t^2}{p_a^2}-1}
  \arctan\left(\sqrt{\frac{4m_t^2}{p_a^2}-1}\right),\nonumber\\
B_b&=&\frac{i\bar\mu^{-2\eps}}{(4\pi)^2}\left[\frac1\eps
  -\ln\pfrac{m_t^2}{\bar\mu^2}+\hat B_b^f\right],\quad
  \hat B_b^f=2\sqrt{\frac{4m_t^2}{p_b^2}-1}
  \arctan\left(\sqrt{\frac{4m_t^2}{p_b^2}-1}\right),\qquad
\end{eqnarray}
and $C_0=i\hat C_0^f/(4\pi)^2$ with
\begin{eqnarray}
\hat C_0^f&=&\frac{-1}\sla\Bigg[
\Li_2\pfrac{m_H^2-p_a^2+p_b^2-\sla}{m_H^2-p_a^2+p_b^2+i\sla x_a}
+\Li_2\pfrac{m_H^2-p_a^2+p_b^2-\sla}{m_H^2-p_a^2+p_b^2-i\sla x_a}
+\strut\nonumber\\&&\strut
-\Li_2\pfrac{m_H^2-p_a^2+p_b^2+\sla}{m_H^2-p_a^2+p_b^2+i\sla x_a}
-\Li_2\pfrac{m_H^2-p_a^2+p_b^2+\sla}{m_H^2-p_a^2+p_b^2-i\sla x_a}
+\strut\nonumber\\&&\strut
+\Li_2\pfrac{m_H^2+p_a^2-p_b^2+\sla}{m_H^2+p_a^2-p_b^2+i\sla x_b}
+\Li_2\pfrac{m_H^2+p_a^2-p_b^2+\sla}{m_H^2+p_a^2-p_b^2-i\sla x_b}
+\strut\nonumber\\&&\strut
-\Li_2\pfrac{m_H^2+p_a^2-p_b^2-\sla}{m_H^2+p_a^2-p_b^2+i\sla x_b}
-\Li_2\pfrac{m_H^2+p_a^2-p_b^2-\sla}{m_H^2+p_a^2-p_b^2-i\sla x_b}
+\strut\nonumber\\&&\strut
+\Li_2\pfrac{m_H^2-p_a^2-p_b^2-\sla}{m_H^2-p_a^2-p_b^2+i\sla x_0}
+\Li_2\pfrac{m_H^2-p_a^2-p_b^2-\sla}{m_H^2-p_a^2-p_b^2-i\sla x_0}
+\strut\nonumber\\&&\strut
-\Li_2\pfrac{m_H^2-p_a^2-p_b^2+\sla}{m_H^2-p_a^2-p_b^2+i\sla x_0}
-\Li_2\pfrac{m_H^2-p_a^2-p_b^2+\sla}{m_H^2-p_a^2-p_b^2-i\sla x_0}\Bigg]
\end{eqnarray}
with $x_0=\sqrt{4m_t^2/m_H^2-1}$, $x_a=\sqrt{4m_t^2/p_a^2-1}$, and
$x_b=\sqrt{4m_t^2/p_b^2-1}$. Numerically, one obtains an integrated rate of
$2.4\times 10^{-11}\GeV$ which is $0.01\%$ of the rate contribution
$\Gamma^{AA}=2.3316(2)\times 10^{-7}\GeV$. In order to have a second example,
we have calculated the one-loop correction to the $HZZ$ vertex via a top quark
loop. The result reads
\begin{eqnarray}
\lefteqn{V_{HZZ}^{\mu\nu}\ =\ \frac{iem_Z}{c_Ws_W}\times
  \frac{m_t^2}{16m_Z^2c_W^2s_W^2}\times\frac{\alpha}{4\pi\lambda^2}
  \bigg(4a_f^2\lambda^2\ln\pfrac{m_t^2}{m_H^2}
  +\strut}\nonumber\\&&\strut\kern-12pt
  +v_f^2\Big\{\hat C_0^f\Big[\lambda\left(4m_H^2p_a^2p_b^2-(m_H^2-p_a^2-p_b^2
  -4m_t^2)\lambda\right)g^{\mu\nu}+\strut\nonumber\\&&\strut\qquad
  +4p_b^2\left(12m_H^2p_a^2p_b^2+(m_H^2+p_a^2+p_b^2+4m_t^2)\lambda\right)
  p_a^\mu p_a^\nu+\strut\nonumber\\&&\strut\qquad
  -2(m_H^2-p_a^2-p_b^2)\left(12m_H^2p_a^2p_b^2+(m_H^2+p_a^2+p_b^2+4m_t^2)
  \lambda\right)p_a^\mu p_b^\nu+\strut\nonumber\\&&\strut\qquad
  -2(m_H^2-p_a^2-p_b^2)\left(12m_H^2p_a^2p_b^2-(m_H^2-p_a^2-p_b^2-4m_t^2)
  \lambda\right)p_b^\mu p_a^\nu+\strut\nonumber\\&&\strut\qquad
  +4p_a^2\left(12m_H^2p_a^2p_b^2+(m_H^2+p_a^2+p_b^2+4m_t^2)\lambda\right)
  p_b^\mu p_b^\nu\Big]+\strut\nonumber\\&&\strut
  +2\hat B_0^f\Big[\lambda\left(m_H^2(m_H^2-p_a^2-p_b^2)-\lambda\right)
  g^{\mu\nu}
  +4p_b^2\left(3m_H^2(m_H^2-p_a^2-p_b^2)-\lambda\right)p_a^\mu p_a^\nu
  +\strut\nonumber\\&&\strut\qquad
  -2(m_H^2-p_a^2-p_b^2)\left(3m_H^2(m_H^2-p_a^2-p_b^2)-\lambda\right)
  p_a^\mu p_b^\nu+\strut\nonumber\\&&\strut\qquad
  -2\left(12m_H^2p_a^2p_b^2+(p_a^2+p_b^2)\lambda\right)p_b^\mu p_a^\nu
  +4p_a^2\left(3m_H^2(m_H^2-p_a^2-p_b^2)-\lambda\right)p_b^\mu p_b^\nu\Big]
  +\strut\nonumber\\&&\strut
  -2\hat B_a^f\Big[p_a^2\lambda(m_H^2-p_a^2+p_b^2)g^{\mu\nu}
  +4p_b^2\left(3p_a^2(m_H^2-p_a^2+p_b^2)+\lambda\right)p_a^\mu p_a^\nu
  +\strut\nonumber\\&&\strut\qquad
  -2(m_H^2-p_a^2-p_b^2)\left(3p_a^2(m_H^2-p_a^2+p_b^2)+\lambda\right)
  p_a^\mu p_b^\nu+\strut\nonumber\\&&\strut\qquad
  -2p_a^2\left(6p_b^2(m_H^2+p_a^2-p_b^2)+\lambda\right)p_b^\mu p_a^\nu
  +4p_a^2\left(3p_a^2(m_H^2-p_a^2+p_b^2)+\lambda\right)p_b^\mu p_b^\nu\Big]
  +\strut\nonumber\\&&\strut
  -2\hat B_b^f\Big[p_b^2\lambda(m_H^2+p_a^2-p_b^2)g^{\mu\nu}
  +4p_b^2\left(3p_b^2(m_H^2+p_a^2-p_b^2)+\lambda\right)p_a^\mu p_a^\nu
  +\strut\nonumber\\&&\strut\qquad
  -2(m_H^2-p_a^2-p_b^2)\left(3p_b^2(m_H^2+p_a^2-p_b^2)+\lambda\right)
  p_a^\mu p_b^\nu+\strut\nonumber\\&&\strut\qquad
  -2p_b^2\left(6p_a^2(m_H^2-p_a^2+p_b^2)+\lambda\right)p_b^\mu p_a^\nu
  +4p_a^2\left(3p_b^2(m_H^2+p_a^2-p_b^2)+\lambda\right)p_b^\mu p_b^\nu\Big]
  \Big\}+\strut\nonumber\\&&\strut\kern-12pt
  +a_f^2\Big\{\hat C_0^f\Big[\lambda\left(4m_H^2p_a^2p_b^2+(m_H^2+p_a^2+p_b^2
  -4m_t^2)\lambda\right)g^{\mu\nu}+\strut\nonumber\\&&\strut\qquad
  +4p_b^2\left(12m_H^2p_a^2p_b^2+(m_H^2+p_a^2+p_b^2+4m_t^2)\lambda\right)
  p_a^\mu p_a^\nu+\strut\nonumber\\&&\strut\qquad
  -2(m_H^2-p_a^2-p_b^2)\left(12m_H^2p_a^2p_b^2+(m_H^2+p_a^2+p_b^2+4m_t^2)
  \lambda\right)p_b^\mu p_a^\nu+\strut\nonumber\\&&\strut\qquad
  -2(m_H^2-p_a^2-p_b^2)\left(12m_H^2p_a^2p_b^2-(m_H^2-p_a^2-p_b^2-4m_t^2)
  \lambda\right)p_a^\mu p_b^\nu+\strut\nonumber\\&&\strut\qquad
  +4p_a^2\left(12m_H^2p_a^2p_b^2+(m_H^2+p_a^2+p_b^2+4m_t^2)\lambda\right)
  p_b^\mu p_b^\nu\Big]+\strut\nonumber\\&&\strut
  +2\hat B_0^f\Big[\lambda\left(m_H^2(m_H^2-p_a^2-p_b^2)-\lambda\right)
  g^{\mu\nu}
  +4p_b^2\left(3m_H^2(m_H^2-p_a^2-p_b^2)-\lambda\right)p_a^\mu p_a^\nu
  +\strut\nonumber\\&&\strut\qquad
  -2(m_H^2-p_a^2-p_b^2)\left(3m_H^2(m_H^2-p_a^2-p_b^2)-\lambda\right)
  p_b^\mu p_a^\nu+\strut\nonumber\\&&\strut\qquad
  -2\left(12m_H^2p_a^2p_b^2+(p_a^2+p_b^2)\lambda\right)p_a^\mu p_b^\nu
  +4p_a^2\left(3m_H^2(m_H^2-p_a^2-p_b^2)-\lambda\right)p_b^\mu p_b^\nu\Big]
  +\strut\nonumber\\&&\strut
  -2\hat B_a^f\Big[\lambda\left(p_a^2(m_H^2-p_a^2+p_b^2)+\lambda\right)
  g^{\mu\nu}
  +4p_b^2\left(3p_a^2(m_H^2-p_a^2+p_b^2)+\lambda\right)p_a^\mu p_a^\nu
  +\strut\nonumber\\&&\strut\qquad
  -2(m_H^2-p_a^2-p_b^2)\left(3p_a^2(m_H^2-p_a^2+p_b^2)+\lambda\right)
  p_b^\mu p_a^\nu+\strut\nonumber\\&&\strut\qquad
  -2p_a^2\left(6p_b^2(m_H^2+p_a^2-p_b^2)+\lambda\right)p_a^\mu p_b^\nu
  +4p_a^2\left(3p_a^2(m_H^2-p_a^2+p_b^2)+\lambda\right)p_b^\mu p_b^\nu\Big]
  +\strut\nonumber\\&&\strut
  -2\hat B_b^f\Big[\lambda\left(p_b^2(m_H^2+p_a^2-p_b^2)+\lambda\right)
  g^{\mu\nu}
  +4p_b^2\left(3p_b^2(m_H^2+p_a^2-p_b^2)+\lambda\right)p_a^\mu p_a^\nu
  +\strut\nonumber\\&&\strut\qquad
  -2(m_H^2-p_a^2-p_b^2)\left(3p_b^2(m_H^2+p_a^2-p_b^2)+\lambda\right)
  p_b^\mu p_a^\nu+\strut\nonumber\\&&\strut\qquad
  -2p_b^2\left(6p_a^2(m_H^2-p_a^2+p_b^2)+\lambda\right)p_a^\mu p_b^\nu
  +4p_a^2\left(3p_b^2(m_H^2+p_a^2-p_b^2)+\lambda\right)p_b^\mu p_b^\nu\Big]
  \Big\}\bigg),\nonumber\\
\end{eqnarray}
where we have used the $\overline{\rm MS}$ scheme to subtract the UV
singularity. Replacing one of the two $HZZ$ vertices in $\Gamma^{AA}$ by this
correction, one obtains an integrated width of $1.9\times 10^{-11}\GeV$ which
is even less than the previous one. With these two estimates at hand, we can
state that class III corrections are negligible as compared to mass effects.

\end{appendix}

\vfill\vspace{24pt}\noindent
{\bf Compliance with Ethical Standards:} the work is in compliance with
ethical standards.\\[12pt]
{\bf Funding:} The work of S.G.\ and M.N.\ has been funded by the European
Regional Development Fund under Grant No.~TK133.\\[12pt]
{\bf Conflict of Interest:} There is no conflict of interest.\\[12pt]
{\bf Ethical Conduct:} ethical conduct is respected.

\end{document}